\documentclass[nofootinbib,pra,twocolumn,showpacs,notitlepage,amsmath,amssymb,superscriptaddress]{revtex4-2} 
\usepackage[colorlinks=true,citecolor=blue,linkcolor=blue,urlcolor=blue]{hyperref}
\usepackage{cleveref}[2012/02/15]
 \crefformat{footnote}{#2\footnotemark[#1]#3}

\usepackage{amsmath,amssymb,bm}

\usepackage{mdframed}
\usepackage{booktabs}

\usepackage{tikz}
\usetikzlibrary{decorations.markings, arrows.meta}
\usetikzlibrary{external}
\usetikzlibrary{decorations.pathreplacing}
\usetikzlibrary{decorations.text}
\usetikzlibrary{decorations.pathmorphing}
\usetikzlibrary{shapes}

\usetikzlibrary{calc}

\definecolor{newblue}{RGB}{112,178,255}
\definecolor{neworange}{RGB}{255,204,112}
\definecolor{blue2}{RGB}{120,0,255}
\definecolor{red2}{RGB}{255,0,120}
\definecolor{green2}{RGB}{0,130,130}

\tikzset{snake it/.style={decorate, decoration={snake,segment length=1mm, amplitude=0.5mm}}}

\definecolor{darkred}{RGB}{245,186,183}
\definecolor{lightred}{RGB}{249,217,215}

\usetikzlibrary{arrows,calc,external,shapes.geometric}

\tikzset{
	>=stealth',
	help lines/.style={dashed, thick},
	important line/.style={thick},
	connection/.style={thick, dotted},
}

\tikzstyle{A}=[circle,draw=red!50,fill=red!20,thick]
\tikzstyle{R}=[circle,draw=blue!50,fill=blue!20,thick]
\tikzstyle{U}=[circle,draw=green!50,fill=green!20,thick]
\tikzstyle{V}=[circle,draw=orange!50,fill=orange!20,thick]

\tikzstyle{bag} = [align=center]

\def\bra#1{\mathinner{\langle{#1}|}}
\def\ket#1{\mathinner{|{#1}\rangle}}

\newcommand{\eeq}{\end{equation}}
\newcommand{\ea}{\end{array}}

\def\eea{\end{eqnarray}}

\def\<{\langle}
\def\>{\rangle}

\usepackage{amsmath}
\usepackage{comment}
\usepackage{amssymb}
\usepackage{amsthm}

\theoremstyle{definition}

\makeatletter
\renewcommand\onecolumngrid{% <<<<<<
\do@columngrid{one}{\@ne}%
\def\set@footnotewidth{\onecolumngrid}% <<<<<<<<<<<<<<<<
\def\footnoterule{\kern-6pt\hrule width 1.5in\kern6pt}
}
\makeatother

\begin{document}
\hfill MIT-CTP/5946

\title{Non-Invertible Interfaces Between Symmetry-Enriched Critical Phases}

\author{Saranesh Prembabu}
\affiliation{Department of Physics, Harvard University, Cambridge, MA 02138, USA}
\author{Shu-Heng Shao}
\affiliation{Center for Theoretical Physics - a Leinweber Institute, Massachusetts Institute of Technology,
Cambridge, MA 02139, USA}
\author{Ruben Verresen}
\affiliation{Pritzker School of Molecular Engineering, University of Chicago, Chicago, IL 60637, USA}

\begin{abstract}
Gapless quantum phases can become distinct when internal symmetries are enforced, in analogy with gapped symmetry-protected topological (SPT) phases. However, this distinction does not always lead to protected edge modes, raising the question of how the bulk-boundary correspondence is generalized to gapless cases. We propose that the spatial \emph{interface} between gapless phases---rather than their boundaries---provides a more robust fingerprint. We show that whenever two 1+1d conformal field theories (CFTs) differ in symmetry charge assignments of local operators or twisted sectors, any symmetry-preserving spatial interface between the theories must flow to a non-invertible defect. We illustrate this general result for different versions of the Ising CFT with $\mathbb{Z}_2 \times \mathbb{Z}_2^T$ symmetry, obtaining a complete classification of allowed conformal interfaces. When the Ising CFTs differ by nonlocal operator charges, the interface hosts 0+1d symmetry-breaking phases with finite-size splittings scaling as $1/L^3$, as well as continuous phase transitions between them. For general gapless phases differing by an SPT entangler, the interfaces between them can be mapped to conformal defects with a certain defect 't Hooft anomaly. This classification also gives implications for higher-dimensional examples, including symmetry-enriched variants of the 2+1d Ising CFT. Our results establish a physical indicator for symmetry-enriched criticality through symmetry-protected interfaces, giving a new handle on the interplay between topology and gapless phases.

\end{abstract}

\maketitle

\tableofcontents

\section{Introduction}

Symmetry-protected topological (SPT) phases are the simplest examples of quantum matter whose characterization requires topological concepts~\cite{Gu_2009, Pollmann_2010, Pollmann_2012,  Turner_2011, Fidkowski_2011, Schuch_2011, Chen_2011, Chen_2011_2, Chen_2013, Chen2012SPT, Else_2014, Wang_2014, Senthil_2015, Kapustin_2015}. In the absence of symmetry considerations, SPTs can be connected by a path of local gapped Hamiltonians, but they become distinct once symmetry is enforced. Recent work has extended this to the gapless setting, leading to the notions of `gapless SPTs' and, more generally, `symmetry-enriched criticality'\footnote{This terminology mirrors that of symmetry-enriched topological order \cite{Barkeshli_2019}, whereby a given universality class is subdivided in the presence of symmetries.} (SEC)~\cite{Kestner11,Grover12,Keselman_2015,Scaffidi_2017,Parker_2018,Verresen_2021,Verresen_2018,verresen2020topologyedgestatessurvive,Thorngren_2021,Duque_2021,Borla_2021,Hidaka_2022,Prembabu_2024,Ma_2022, Wen_2023, wen2023classification11dgaplesssymmetry,Li_2024,Lei24,Yu24,Ando24,Bhardwaj:2024qrf,qi2025symmetrytacoequivalencesgapped,Linhao25,Calderon25,wen2025topologicalholography21dgapped,Antinucci25,Rey_2025, prembabu2025multicriticalitypurelygaplessspt}. The basic idea is similar: given two lattice Hamiltonians that can be connected by a path of local Hamiltonians flowing to the \emph{same} universality class at each point, we say they belong to \emph{distinct} symmetry-enriched classes if no such path exists when enforcing a specified microscopic \emph{symmetry}. Gapped SPTs arise as the special case where the chosen universality class is the trivial (or `empty') one.

Despite this appealing parallel, key questions remain. One concerns the physical fingerprints of gapless SPTs: while gapped SPTs have clear boundary signatures, gapless examples without edge degeneracies are now known \cite{Prembabu_2024},
opening up the search for universal physical indicators.
A second challenge is the use of the path-based definition itself.
Existing diagnostics provide obstructions to symmetric paths---such as mismatched symmetry charges of low-energy scaling operators \cite{Verresen_2021}. 
The converse direction is much less obvious, unlike in the gapped case.

To further highlight the differences between the gapped and gapless cases, we focus on internal symmetries: on-site unitary or time-reversal symmetry. For gapped SPTs defined on a tensor product Hilbert space, there is a canonically trivial phase, and all nontrivial phases carry edge modes and are classified by their twisted partition functions. In the gapless setting, by contrast, some SECs are distinguished by the charges of \emph{local} operators (see Sec.~III); while it is exciting to have such a new class of options which do not exist in the gapped case, they are not characterized by edge modes. Other SECs differ through the charges of \emph{nonlocal} scaling operators; these are often termed “gapless SPTs.’’ Early examples all exhibited protected edge degeneracies \cite{Keselman_2015, Scaffidi_2017, Verresen_2018, verresen2020topologyedgestatessurvive, Thorngren_2021, Duque_2021, wang2023stabilityfinestructuresymmetryenriched, chou2025ptsymmetryenrichednonunitarycriticality,Zhou_2025}, mirroring the gapped case, but a recent counterexample~\cite{Prembabu_2024} admits boundary conditions with a unique ground state.\footnote{This example has not yet been studied from an entanglement-based perspective, although for other gapless SPTs see Ref.~\cite{Hidaka_2022,PhysRevLett.129.210601, Choi_2024,Zhong_2025, banerjee2025entanglementspectrumgaplesstopological}.} This sharpens the question of what is the appropriate physical diagnostic that detects these distinct gapless phases?

In this work we address these issues by focusing on \emph{interfaces} between symmetry-enriched critical systems. This perspective is motivated by the gapped case, where an edge can be viewed as a spatial interface to the trivial SPT, and two gapped SPTs lie in the same phase \emph{if and only if} their interface carries no protected modes (for internal symmetries). This suggests that interfaces---not edges---might be the fundamental objects for diagnosing distinctions between critical chains with internal symmetry $G$. 
In particular, this perspective has been highlighted in Refs.~\onlinecite{Fechisin:2023odt,Seifnashri:2024dsd} in the context of gapped SPT phases protected by non-invertible symmetries, where there may not be a canonically trivial phase.

\begin{figure}[t]
\centering

\begin{tikzpicture}[baseline, line cap=round, line join=round]

% --- colors ---
\colorlet{defc}{purple!75!black}  % D? + guide arrows
\colorlet{Tzeroc}{blue!70!black}  % T_0 side
\colorlet{Tonec}{red!70!black}    % T_1 side

  % Parameters
  \def\xL{-4.1}
  \def\xR{4.1}
  \def\y{0}
  \def\boxW{0.78}
  \def\boxH{0.7}
  \def\yarr{-1.1}
  \def\arrowLen{1.15}
  \def\arrowDx{2.95}
  % four icon centers (tweak these to change spacing)
    \def\xSSB{-3.10}
    \def\xFAC{-1.05}
    \def\xCONF{ 1.00}
    \def\xNIT{ 3.10}
    
  \colorlet{defc}{violet!70!black}

  % Horizontal line (broken at D?)
  \draw[thick, Tzeroc] (\xL,\y) -- (-\boxW/2,\y);
  \draw[thick, Tonec] (\boxW/2,\y) -- (\xR,\y);

  % Box for D?
  \node[draw=defc, text=defc, minimum width=\boxW cm, minimum height=\boxH cm, fill=white, inner sep=2pt, thick] (dbox) at (0,\y) {\Large $\mathcal{D}$};

  % T0 and T1 inside the line
  \node[above] at (-2.4,0) {\large $\mathcal{T}_0$};
  \node[above] at (2.4,0) {\large $\mathcal{T}_1$};
  \node[below] at (-2.4,0) {\tiny e.g. $\phi(z,\overline z) \sim Y_j$};
  \node[below] at (2.4,0) {\tiny e.g. $\phi(z,\overline z) \sim Z_j$};

  % --- curved, fanned arrows from the box to the four icons ---
\tikzset{fan/.style={thin, defc, ->, shorten >=1.5pt}} % tweak shorten if needed
\def\S{(dbox.south)}  % start at bottom of the D box

\draw[fan] \S to[out=-150, in=70, looseness=0.78] (\xSSB ,\yarr); % SSB (far left)
\draw[fan] \S to[out=-120, in=60, looseness=1.03] (\xFAC ,\yarr); % factorizing (left)
\draw[fan] \S to[out= -60, in= 120, looseness=0.98] (\xCONF,\yarr); % conformal (center)
\draw[fan] \S to[out= -30, in= 110, looseness=0.78] (\xNIT ,\yarr); % non-invertible topo (right)

\newcommand{\ScatIcon}[3]{%
  \begin{scope}[shift={(#1,#2)}, scale=#3]
    \def\H{0.62}      % height of vertical bar
    \def\L{0.90}      % arrow length
    \def\m{0.30}      % slope for incoming & transmitted (dy/dx)
    \def\amp{0.03cm}  % wave amplitude
    \def\seg{0.1cm}  % wavelength along the path

    % interface
    \draw[line width=1.0pt, line cap=round, violet] (0,-\H/2) -- (0,\H/2);

    % components for slope m
    \pgfmathsetmacro{\dx}{\L/sqrt(1+\m*\m)}
    \pgfmathsetmacro{\dy}{\m*\dx}

    % nice wave styling
    \tikzset{wave/.style={
      yellow!70!black, line width=0.5pt, ->,
      decorate, decoration={snake, amplitude=\amp, segment length=\seg,
                            pre length=0.02cm, post length=0.06cm}
    }}
    % incoming, transmitted, reflected (same length L)
    \draw[->,wave] (-\dx,-\dy) -- (-\dx*0.1,-\dy*0.1);   % incoming
    \draw[->,wave] (0,0) -- (\dx*0.71,\dy*0.71);     % transmitted
    \draw[->,wave] (0,0) -- (-\dx*0.71,\dy*0.71);    % reflected (slope = -m)
    % want reflected down-left instead? use (-- -\dx,-\dy)
  \end{scope}%
}

\newcommand{\ScatIconReflect}[3]{%
  \begin{scope}[shift={(#1,#2)}, scale=#3]
    \def\H{0.62}      % height of vertical bar
    \def\L{0.90}      % arrow length
    \def\m{0.30}      % slope for incoming & transmitted (dy/dx)
    \def\amp{0.03cm}  % wave amplitude
    \def\seg{0.1cm}  % wavelength along the path

    % interface
    \draw[line width=1.0pt, line cap=round, violet] (0,-\H/2) -- (0,\H/2);

    % components for slope m
    \pgfmathsetmacro{\dx}{\L/sqrt(1+\m*\m)}
    \pgfmathsetmacro{\dy}{\m*\dx}

    % nice wave styling
    \tikzset{wave/.style={
      yellow!70!black, line width=0.5pt, ->,
      decorate, decoration={snake, amplitude=\amp, segment length=\seg,
                            pre length=0.02cm, post length=0.06cm}
    }}
    % incoming, transmitted, reflected (same length L)
    \draw[->,wave] (-\dx,-\dy) -- (-\dx*0.1,-\dy*0.1);   % incoming
    %\draw[->,wave] (0,0) -- (\dx,\dy);     % transmitted
    \draw[->,wave] (0,0) -- (-\dx,\dy);    % reflected (slope = -m)
    % want reflected down-left instead? use (-- -\dx,-\dy)
  \end{scope}%
}

\newcommand{\ScatIconTransmit}[3]{%
  \begin{scope}[shift={(#1,#2)}, scale=#3]
    \def\H{0.62}      % height of vertical bar
    \def\L{0.90}      % arrow length
    \def\m{0.30}      % slope for incoming & transmitted (dy/dx)
    \def\amp{0.03cm}  % wave amplitude
    \def\seg{0.1cm}  % wavelength along the path

    % interface
    \draw[line width=1.0pt, line cap=round, violet] (0,-\H/2) -- (0,\H/2);

    % components for slope m
    \pgfmathsetmacro{\dx}{\L/sqrt(1+\m*\m)}
    \pgfmathsetmacro{\dy}{\m*\dx}

    % nice wave styling
    \tikzset{wave/.style={
      yellow!70!black, line width=0.5pt, ->,
      decorate, decoration={snake, amplitude=\amp, segment length=\seg,
                            pre length=0.02cm, post length=0.06cm}
    }}
    % incoming, transmitted, reflected (same length L)
    \draw[->,wave] (-\dx,-\dy) -- (-\dx*0.1,-\dy*0.1);   % incoming
    \draw[->,wave] (0,0.1) -- (\dx*0.71,\dy*0.71+0.1);     % transmitted
    \draw[->,wave] (0,-0.1) -- (\dx*0.71,\dy*0.71-0.1);     % transmitted
    %\draw[->,wave] (0,0) -- (-\dx,\dy);    % reflected (slope = -m)
    % want reflected down-left instead? use (-- -\dx,-\dy)
  \end{scope}%
}

% SSB icon: interface with two short vertical arrows (edge modes)
% SSB = (interface + up-right arrow)  ⊕  (interface + down-left arrow)
\newcommand{\ScatIconSSB}[3]{%
  \begin{scope}[shift={(#1,#2)}, scale=#3]
    % knobs
    \def\H{0.62}      % bar height
    \def\sep{0.62}    % half separation between the two interface icons
    \def\L{0.57}      % arrow length
    \def\angUR{50}    % degrees: up-right tilt (near vertical)
    \def\angDL{180+50}   % degrees: down-left tilt (= -105)

    % colors (use yours if defined elsewhere)
    \colorlet{barc}{violet}
    \colorlet{edgc}{black!70!black}

    % left icon: bar + up-right arrow
    \draw[line width=1.0pt, line cap=round, barc] (-\sep,-\H/2) -- (-\sep,\H/2);
    \draw[->, line width=0.6pt, edgc] (-\sep,0) -- ++(\angUR:\L);

    % middle plus
    \node at (0,0) {$\oplus$};

    % right icon: bar + down-left arrow
    \draw[line width=1.0pt, line cap=round, barc] (\sep,-\H/2) -- (\sep,\H/2);
    \draw[->, line width=0.6pt, edgc] (\sep,0) -- ++(\angDL:\L);
  \end{scope}%
}

% place four icons at the arrow tips
\ScatIconSSB     {\xSSB }{\yarr-0.5}{0.62}   % SSB
\ScatIconReflect {\xFAC }{\yarr-0.5}{0.62}   % factorizing (R only)
\ScatIcon        {\xCONF}{\yarr-0.5}{0.62}   % conformal (R+T)
\ScatIconTransmit{\xNIT }{\yarr-0.5}{0.62}   % non-invertible topological (T only)

% labels under the four scattering icons
\def\labdy{0.55}
\tikzset{scatlabel/.style={font=\footnotesize, anchor=north, align=center}}
\node[scatlabel] at (\xSSB , \yarr-0.5-\labdy) {SSB};
\node[scatlabel] at (\xFAC , \yarr-0.5-\labdy) {Factorizing};
\node[scatlabel] at (\xCONF, \yarr-0.5-\labdy) {Conformal};
\node[scatlabel] at (\xNIT , \yarr-0.5-\labdy) {Non-invertible\\Topological};

\node[] at ([xshift=8pt,yshift=5pt]{\arrowDx+0.5,\yarr-0.5-\labdy+0.1}) { \scriptsize $\langle \mathcal{D} \rangle>1$};

\node[anchor=north west, font=\bfseries\small, xshift=1pt, yshift=-1pt]
  at (current bounding box.north west) {(a)};

\end{tikzpicture}

\resizebox{0.98\columnwidth}{!}{%
\begin{tikzpicture}[x=1cm,y=1cm,line cap=round,line join=round,>=stealth]
  % colors
  \colorlet{defc}{violet!90!black}  % distinct from red
  \colorlet{gc}{blue!75!black}      % group-side lines (rho_0)
  \colorlet{Ac}{red!75!black}       % A-lines (rho_1)

  % styles (PRL column)
  \tikzset{
    defline/.style={defc, line width=1.2pt},
    gline/.style={gc,   line width=1.2pt},
    Aline/.style={Ac,   line width=1.2pt},
    blab/.style={text=Ac, font=\footnotesize}, % rho_1 labels (red)
    glab/.style={text=gc, font=\footnotesize}, % rho_0 labels (blue)
    dlab/.style={text=defc, font=\footnotesize}
  }

  % compact geometry
  \def\sep{4.6}
  \def\H{0.95}
  \def\R{1.55}
  \def\yoff{0.42}
  \def\xsplit{\R-0.75}

  % ---------------- LEFT PANEL (LHS) ----------------
  \begin{scope}[shift={(-\sep/2,0)}]
    % background tints (blue left, red right)
    \fill[blue!6] (-\R,-\H) rectangle (0,\H);
    \fill[red!6]  (0,-\H)  rectangle (\R,\H);

    % vertical defect
    \draw[defline] (0,-\H) -- (0,\H);
    \node[dlab, anchor=north west] at (0.05,-\H) {$\mathcal{D}$};

    % blue rho_0(gh) on the line itself
    \draw[gline] (-\R,0) -- node[glab, pos=0.35, above] {$\rho_0(gh)$} (0,0);

    % red trunk rho_1(gh) on the trunk
    \draw[Aline] (0,0) -- node[blab, pos=0.6, above] {$\rho_1(gh)$} (\xsplit,0);

    \node[glab] at (-0.4,-0.6){$\mathcal T_0$};
    \node[blab] at (0.4,-0.6){$\mathcal T_1$};

    % upper red leg: curved, label on the curve
    \draw[Aline] (\xsplit,0)
      .. controls (\xsplit+0.35, 0.28) and (\R-0.35, \yoff) ..
      node[blab, pos=0.62, above] {$\rho_1(g)$}
      (\R,\yoff);

    % lower red leg: curved, label on the curve
    \draw[Aline] (\xsplit,0)
      .. controls (\xsplit+0.35,-0.28) and (\R-0.35,-\yoff) ..
      node[blab, pos=0.62, below] {$\rho_1(h)$}
      (\R,-\yoff);

    % ---- phase annotations (LHS) ----
    % at the red split junction (\xsplit,0): alpha_1
    %\node[blab, right=+3pt, yshift=2pt] at (\xsplit,0) { $e^{i\alpha_1}$};
  \end{scope}

  % equals sign
  \node[font=\small] at (0,0) {$=$};

  % ---------------- RIGHT PANEL (RHS) ----------------
  \begin{scope}[shift={( \sep/2,0)}]
    % background tints
    \fill[blue!6] (-\R,-\H) rectangle (0,\H);
    \fill[red!6]  (0,-\H)  rectangle (\R,\H);

    % vertical defect
    \draw[defline] (0,-\H) -- (0,\H);
    \node[dlab, anchor=north west] at (0.05,-\H) {$\mathcal{D}$};

    % red straight horizontals with labels ON the lines
    \draw[Aline] (0,\yoff) -- node[blab, pos=0.58, above] {$\rho_1(g)$} (\R,\yoff);
    \draw[Aline] (0,-\yoff) -- node[blab, pos=0.58, below] {$\rho_1(h)$} (\R,-\yoff);

    % blue curves fusing to rho_0(gh); labels ON the curves
    \coordinate (M) at (-0.75,0);
    \draw[gline] (0,\yoff)
      .. controls (-0.35,\yoff) and (-0.55,0.30) ..
      node[glab, pos=0.45, above] {$\rho_0(g)$}
      (M);
    \draw[gline] (0,-\yoff)
      .. controls (-0.35,-\yoff) and (-0.55,-0.30) ..
      node[glab, pos=0.45, below] {$\rho_0(h)$}
      (M);
    \draw[gline] (-\R,0) -- node[glab, pos=0.55, above] {$\rho_0(gh)$} (M);

  \end{scope}
  \path[overlay] node[anchor=north west, font=\bfseries\small,
                    xshift=1pt, yshift=12pt]
  at (current bounding box.north west) {(b)};
\end{tikzpicture}%

}
\caption{\textbf{RG fate of $G$-symmetric interfaces $\mathcal D$ between symmetry-enriched criticalities.} (a) ``Symmetry-enriched criticalities'' $\mathcal T_0$ and $\mathcal T_1$ share the same low-energy conformal field theory (CFT) but differ by how the global symmetry $G$ acts on (local or nonlocal) CFT operators. We classify the universal RG outcomes of any spatial interface $\mathcal D$ that preserves $G$. Bulk symmetry enrichment leaves concrete physically-observable interface signatures. In particular, symmetry forbids an invertible defect; instead $\mathcal D$ may flow to a \textit{non-invertible} topological defect (with non-invertible fusion and quantum dimension $\langle \mathcal{D} \rangle>1$), to a partially or totally reflecting (factorizing) interface, or to a degenerate interface spontaneously breaking $G$. (b) In the IR, we say an interface $\mathcal D$ is $G$-symmetric when the UV network of $G$-defect lines and their junctions remains topological in its presence, i.e., it can slide across $\mathcal D$ without changing correlators or partition-function evaluations. Crucially, the UV $G$-defect lines map to IR defects and/or three-way junction phases in \textit{different} ways in $\mathcal{T}_0$ vs. $\mathcal{T}_1$. }
\label{fig:G-symmetric-defect}
\end{figure}
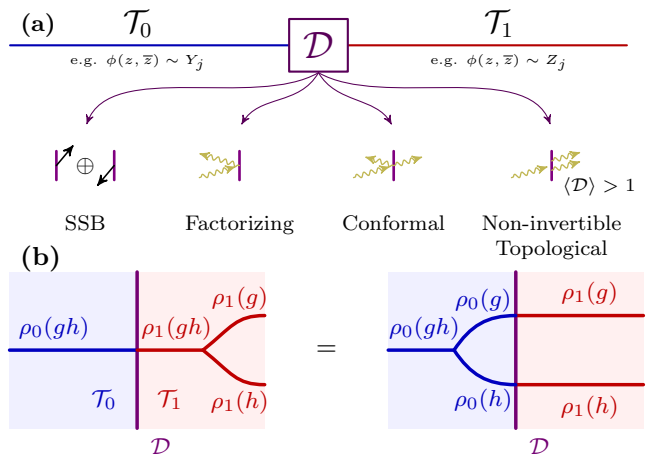

Our first result is that interfaces provide a physical diagnostic of SECs. Whenever two critical theories differ in the symmetry charges of local scaling operators or of twisted-sector operators, any symmetric interface between them necessarily flows to a non-transparent defect. 
In particular, the interface cannot be an \textit{invertible} topological defect (Fig.~\ref{fig:G-symmetric-defect}(a)). 
The non-invertible nature of these defects has directly observable signatures that are absent for invertible defects. For example:
\begin{enumerate}
  \item[(i)] Point operators on the interface have distinct correlation functions and universal scaling dimensions not seen in the bulk, corresponding to different defect operators only present at the interface. 
  \item[(ii)] The low-energy finite-size scaling of the energy spectrum of a system on a ring with two distinct SECs on two halves of the ring differs from the spectrum of a uniform system. (This is related to the previous point by the operator-state correspondence.) 
  \item[(iii)] Wave packets incident on the interface may scatter or transform non-trivially (Fig.~\ref{fig:G-symmetric-defect} (a)) and energy transmitted across the interface necessarily generates entropy \cite{Quella_2007, Bachas_2008}.
\end{enumerate}

Interfaces are therefore at least as sensitive a diagnostic as the charge assignments of local and nonlocal scaling operators, while being more directly tied to physical observables.

A second advantage is that interfaces have a more natural connection to the path-based definition. A spatial interface is itself a kind of interpolation between theories and can often be analyzed more directly than a Hamiltonian path, as it reduces to the study of a 0+1d feature (or more generally a lower-dimensional feature). Although our goal is not to provide a rigorous equivalence of the path- and interface-based notions, we explain their intuitive connection, which could form the basis for future work.

In addition to our general results, we illustrate these ideas using examples based on the Ising conformal field theory (CFT) with $\mathbb{Z}_2 \times \mathbb{Z}_2^T$ symmetry, covering both local and nonlocal charge assignments. In every case where the theories are symmetry-inequivalent, we find no symmetric invertible topological defects. Instead, the interface phase diagram contains non-invertible topological interfaces (such as the Kramers–Wannier defect) and  conformal interfaces, sometimes with protected ground-state degeneracies. 
These examples show that interfaces not only detect SEC distinctions but also host a rich variety of physical phenomena. Moreover, we discuss how these ideas generalize to higher dimensions.

\section{Non-invertible defects and vanishing correlators \label{sec:noninvertible}}

Let $G$ be some ultraviolet (UV) global symmetry of interest; in this work we focus on the case of internal and time-reversal symmetries. 
Consider a $G$-symmetric  interface \(\mathcal{D}\) separating the theories \(\mathcal{T}_0\) and \(\mathcal{T}_1\).
If we ignore the $G$ symmetry action, then \(\mathcal{T}_0\) and \(\mathcal{T}_1\) share the same underlying CFT data, i.e., they share the same spectrum of local operators and OPE coefficients. 
Hence at low energies \(\mathcal{D}\) can be viewed as a conformal defect of this CFT. 

A special class of  conformal defects are topological defects, which commute with the stress-energy tensor \cite{Petkova_2001}. 
See, for example, Refs. \onlinecite{Bhardwaj:2017xup,Chang:2018iay,Schafer-Nameki:2023jdn,Shao:2023gho,Carqueville:2023jhb} for recent discussions of topological defects in 1+1d CFTs. 
A topological defect $\cal D$ is called invertible  if there is an inverse defect \(\mathcal{D}^{-1}\) with which it fuses to the identity; we naturally say an interface is invertible if it flows to an invertible defect. 
Since the fusion of conformal defects are subject to short-distance singularity \cite{Bachas_2008}, invertible defects are necessarily topological.\footnote{Indeed, any defect which scatters energy cannot be completely inverted by another defect.} 
Our first main result will be the following:

\medskip

\noindent
\fbox{%
\parbox{\dimexpr\linewidth-2\fboxsep-2\fboxrule}{%
\textbf{Claim.} If there exists an interface \(\mathcal{D}\) between \(\mathcal{T}_0\) and \(\mathcal{T}_1\) which is $G$-symmetric and \emph{invertible}, then \(\mathcal{T}_0\) and \(\mathcal{T}_1\) have the \emph{same} symmetry-charge assignments for all operators, local and nonlocal.%
}%
}

\medskip

Consequently, any admissible interface between two \emph{distinct} symmetry-enriched criticalities must be non-invertible. 
There is no $G$-symmetric local deformation that can trivialize these signatures into those of an invertible defect.

We give two complementary arguments for the claim, which also provide additional insights about defects between symmetry-enriched criticalities from both IR and UV perspectives of symmetry actions. 
The first develops our picture for unitary internal symmetries in bosonic CFTs,
while the second applies more broadly to bosonic or fermionic theories including time-reversal symmetries.

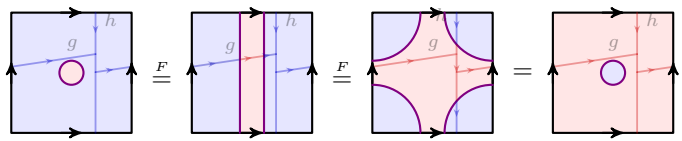
\begin{figure}
\centering

\begin{tikzpicture}[line cap=round, line join=round, scale=0.8]

% --- global params (same as yours) ---
\def\rA{0.2}
\def\rB{0.2}
\def\rC{0.2}
\def\rD{0.2}
\def\width{1.0}
\def\length{1.0}
\def\radC{0.8}
\def\radD{0.2}

\def\dx{3.0}
\def\eqshift{1.5}

\def\hx{0.40}
\def\gy{0.10}
\def\dy{0.15}

\tikzset{
  gaugeedge/.style={black, opacity=0.35, line width=0.7pt},
  glabel/.style={black, opacity=0.35, font=\scriptsize},
  midarrow/.style={
    postaction={decorate},
    decoration={markings, mark=at position 0.50 with {\arrow{Stealth[length=3pt]}}}
  }
}

% ---------- helper to place torus arrows (used in every panel) ----------
\newcommand{\torusarrows}{%
  \draw[->, thick] (-0.175,\length) -- ++(0.35,0);
  \draw[->, thick] (\width,-0.175) -- ++(0,0.35);
  \draw[->, thick] (-0.175,-\length) -- ++(0.35,0);
  \draw[->, thick] (-\width,-0.175) -- ++(0,0.35);
}

% ===================== 1) first square =====================
\begin{scope}[shift={(0,0)}]
  % backgrounds + frame
  \fill[blue!10] (-\width,-\length) rectangle (\width,\length);
  \fill[red!10] (0,0) circle (\rA);
  \draw[thick] (-\width,-\length) rectangle (\width,\length);
  \draw[thick,violet] (0,0) circle (\rA);

  % --- gauge network (panel-specific colors) ---
  \colorlet{gcolA}{blue!80!black}
  \colorlet{hcolA}{blue!80!black}

  \coordinate (A1) at (-\width,\gy);
  \coordinate (B1) at (\hx,{\gy + \dy*(\width+\hx)/\width});
  \coordinate (C1) at (\hx,{\gy - \dy*((\width-\hx)/\width)});
  \coordinate (D1) at (\width,\gy);
  \coordinate (E1) at (\hx,\length);
  \coordinate (F1) at (\hx,-\length);

  \draw[gaugeedge,midarrow,draw=gcolA] (A1)--(B1) node[glabel, pos=0.55, anchor=south west] {$g$};
  \draw[gaugeedge,midarrow, draw=gcolA] (C1)--(D1);
  \draw[gaugeedge,midarrow,draw=hcolA] (E1)--(B1) node[glabel, pos=0.55, anchor=south west] {$h$};
  \draw[gaugeedge, draw=hcolA] (B1)--(C1);
  \draw[gaugeedge, draw=hcolA] (C1)--(F1);

  % torus arrows
  \torusarrows
\end{scope}

% =_F
\node at (\eqshift,0) {\normalsize $\stackrel{\scriptscriptstyle F}{=}$};

% ===================== 2) second square =====================
\begin{scope}[shift={(\dx,0)}]
  \fill[blue!10] (-\width,-\length) rectangle (\width,\length);
  \fill[red!10] (-\rB,-1) rectangle (\rB,1);
  \draw[thick] (-\width,-\length) rectangle (\width,\length);
  \draw[thick,violet] (-\rB,-1) -- (-\rB,1);
  \draw[thick,violet] (\rB,-1) -- (\rB,1);

  % --- gauge network (panel-specific colors) ---
  \colorlet{gcolB}{blue!80!black}
  \colorlet{hcolB}{blue!80!black}

  \coordinate (A2) at (-\width,\gy);
  \coordinate (B2) at (\hx,{\gy + \dy*(\width+\hx)/\width});
  \coordinate (C2) at (\hx,{\gy - \dy*((\width-\hx)/\width)});
  \coordinate (D2) at (\width,\gy);
  \coordinate (E2) at (\hx,\length);
  \coordinate (F2) at (\hx,-\length);

  % Define points for color transition
    \path let \p1 = (A2), \p2 = (B2) in
  coordinate (Gleft) at (-\rB, {\y1 + (\y2-\y1)*((-\rB-\x1)/(\x2-\x1))}) % intersection at x=-\rB
  coordinate (Gright) at (\rB, {\y1 + (\y2-\y1)*((\rB-\x1)/(\x2-\x1))}); % intersection at x=+\rB
    
    % g edge: blue left, red center, blue right
   \draw[gaugeedge,draw=blue!70!black,midarrow] (-1.0,0.10) -- (-0.2,0.22) node[glabel, pos=0.5, anchor=south west] {$g$};;
\draw[gaugeedge,draw=red!80!black,midarrow] (-0.2,0.22) -- (0.2,0.28);
\draw[gaugeedge,draw=blue!70!black,midarrow] (0.2,0.28) -- (0.40,0.31);
  \draw[gaugeedge,midarrow,draw=gcolB] (C2)--(D2);
  \draw[gaugeedge,midarrow,draw=hcolB] (E2)--(B2) node[glabel, pos=0.55, anchor=south west] {$h$};
  \draw[gaugeedge,draw=hcolB] (B2)--(C2);
  \draw[gaugeedge,draw=hcolB] (C2)--(F2);

  \torusarrows
\end{scope}

% =_F
\node at (\dx+\eqshift,0) {\normalsize $\stackrel{\scriptscriptstyle F}{=}$};

% ===================== 3) third square =====================
\begin{scope}[shift={(2*\dx,0)}]
  \fill[red!10] (-\width,-\length) rectangle (\width,\length);
  \fill[blue!10] (1,1) -- (1,1-\radC) arc[start angle=270,end angle=180,radius=\radC] -- cycle;
  \fill[blue!10] (-1,1) -- (-1+\radC,1) arc[start angle=0,end angle=-90,radius=\radC] -- cycle;
  \fill[blue!10] (-1,-1) -- (-1,-1+\radC) arc[start angle=90,end angle=0,radius=\radC] -- cycle;
  \fill[blue!10] (1,-1) -- (1-\radC,-1) arc[start angle=180,end angle=90,radius=\radC] -- cycle;
  \draw[thick] (-\width,-\length) rectangle (\width,\length);
  \draw[thick,violet] (1,1-\radC) arc[start angle=270,end angle=180,radius=\radC];
  \draw[thick,violet] (-1+\radC,1) arc[start angle=0,end angle=-90,radius=\radC];
  \draw[thick,violet] (-1,-1+\radC) arc[start angle=90,end angle=0,radius=\radC];
  \draw[thick,violet] (1-\radC,-1) arc[start angle=180,end angle=90,radius=\radC];

% Define yRedBottom and yRedTop *numerically* with \pgfmathsetmacro
\pgfmathsetmacro{\yRedBottom}{-\length + sqrt(\radC*\radC - (\width-\hx)*(\width-\hx))}
\pgfmathsetmacro{\yRedTop}{\length - sqrt(\radC*\radC - (\width-\hx)*(\width-\hx))}

  % --- gauge network (panel-specific colors) ---
  \colorlet{gcolC}{red!80!black}
  \colorlet{hcolC}{blue!80!black}

  \coordinate (A3) at (-\width,\gy);
  \coordinate (B3) at (\hx,{\gy + \dy*(\width+\hx)/\width});
  \coordinate (C3) at (\hx,{\gy - \dy*((\width-\hx)/\width)});
  \coordinate (D3) at (\width,\gy);
  \coordinate (E3) at (\hx,\length);
  \coordinate (F3) at (\hx,-\length);

    \draw[gaugeedge,midarrow,draw=gcolC] (A3)--(B3) node[glabel, pos=0.55, anchor=south west] {$g$};
    \draw[gaugeedge,midarrow,draw=gcolC] (C3)--(D3);
% Blue segment: from top down to red start
\draw[gaugeedge,midarrow,draw=hcolC] (\hx,\length) -- (\hx,\yRedTop) node[glabel, pos=0.55, anchor=south east] {$h$};
% Red segment: through the middle
\draw[gaugeedge,midarrow,draw=gcolC] (\hx,\yRedTop) -- (\hx,\yRedBottom);
% Blue segment: from red end to bottom
\draw[gaugeedge,midarrow,draw=hcolC] (\hx,\yRedBottom) -- (\hx,-\length);

  \torusarrows
\end{scope}

% =
\node at (2*\dx+\eqshift,0) {\normalsize $=$};

% ===================== 4) fourth square =====================
\begin{scope}[shift={(3*\dx,0)}]
  \fill[red!10] (-\width,-\length) rectangle (\width,\length);
  \fill[blue!10] (0,0) circle (\radD);
  \draw[thick] (-\width,-\length) rectangle (\width,\length);
  \draw[thick,violet] (0,0) circle (\radD);

  % --- gauge network (panel-specific colors) ---
  \colorlet{gcolD}{red!80!black}
  \colorlet{hcolD}{red!80!black}

  \coordinate (A4) at (-\width,\gy);
  \coordinate (B4) at (\hx,{\gy + \dy*(\width+\hx)/\width});
  \coordinate (C4) at (\hx,{\gy - \dy*((\width-\hx)/\width)});
  \coordinate (D4) at (\width,\gy);
  \coordinate (E4) at (\hx,\length);
  \coordinate (F4) at (\hx,-\length);

  \draw[gaugeedge,midarrow,draw=gcolD] (A4)--(B4) node[glabel, pos=0.55, anchor=south west] {$g$};
  \draw[gaugeedge,midarrow,draw=gcolD] (C4)--(D4);
  \draw[gaugeedge,midarrow,draw=hcolD] (E4)--(B4) node[glabel, pos=0.55, anchor=south west] {$h$};
  \draw[gaugeedge,draw=hcolD] (B4)--(C4);
  \draw[gaugeedge,draw=hcolD] (C4)--(F4);

  \torusarrows
\end{scope}

\end{tikzpicture}

\caption{\textbf{Sweeping an invertible interface on a torus does not change its $G$-twisted partition function}. 
 If theories $\mathcal{T}_0$ (blue) and $\mathcal{T}_1$ (red) admit a $G$-symmetric \textit{invertible} topological interface, then they have identical twisted partition functions and thus are \textit{not} distinct symmetry enriched criticalities. Starting from $\mathcal{T}_0$ one can nucleate the invertible interface (left), apply two $F$-symbol moves to sweep it over the whole torus (middle), and contract it again (right), resulting in $\mathcal{T}_1$ and no changes to the partition function in the process. Due to $G$-symmetry, the same equalities hold in the presence of a background $G$-network (grey) defining the $(g,h)$ twisted partition functions. 
 }
\label{fig:sweeping}
\end{figure}

\subsection{Sweeping $G$-symmetric interfaces in the IR \label{subsec:sweep}}

Before proceeding, we need to clarify what we mean by a $G$-symmetric interface in both the UV and IR.
We start from a UV symmetry group $G$, generated by a set of lattice operators that we assume to be onsite for simplicity.
Consider two Hamiltonians, $H_0$ and $H_1$, each defined on a closed chain and both commuting with the $G$-symmetry operators.
We then construct an interface Hamiltonian that coincides with $H_0$ on one half of the chain and with $H_1$ on the other half. 
The local terms at the two interfaces will look different from those of $H_0$ and $H_1$, but we choose them in such a way that the entire interface Hamiltonian commutes with the original $G$-symmetry operators defined on the closed chain, without any modification. 
Such an interface, denoted by $\cal D$, is called $G$-symmetric. 
This setup has the advantage that the symmetry operators stay the same in systems with or without the interfaces. 
See Appendix D of Ref. \onlinecite{Seifnashri:2024dsd} and Appendix \ref{app:gappedSPT} of this paper for an illustration of this setup. 

In the IR, the Hamiltonians $H_0$ and $H_1$ flow to the CFTs ${\cal T}_0$ and ${\cal T}_1$, respectively. 
The same UV symmetry group $G$, however, generally maps to distinct subgroups of the IR symmetry.
We denote the corresponding group homomorphisms\footnote{We note that we do not presume $\rho(G)$ to act faithfully on the local operators in the IR CFT. For instance, $\rho(G)$ is not faithful in the cases of gapped SPTs and intrinsically gapless SPTs \cite{Thorngren_2021}.\label{footnote:faithful}} as $\rho_0$ and $\rho_1$. 
For simplicity, let us restrict our attention to internal unitary symmetries of bosonic CFTs in the following discussion.
Since the UV symmetry operator commutes with the interface Hamiltonian, there must exist topological junctions between the topological line defects associated with the $G$-symmetry and the interface $\cal D$ in the IR CFT. 
Moreover, since the UV symmetry is assumed to be onsite, no additional phase factors arise when a topological $G$-defect line is moved across the interface $\cal D$; see Fig.~\ref{fig:G-symmetric-defect}(b).
(We will explain how $G$-symmetric \textit{interfaces} between distinct theories are related to $G$-symmetric \textit{defects} and their defect anomalies within a single theory  in Section \ref{sec:projective}.)

Therefore, if the interface is topological and $G$-symmetric, it can be swept continuously through any fixed background of $G$-defect lines without changing any correlations or partition functions.
Furthermore, if the defect is \textit{invertible} one can sweep it entirely over the torus, proving that $\mathcal{T}_0$ and $\mathcal{T}_1$ actually have the same twisted partition functions and symmetry charge assignments.

To perform this sweeping, one can nucleate a contractible loop of $\mathcal{D}$ in a region of $\mathcal{T}_0$. Then one can apply two $F$-symbol moves to sweep the $\mathcal{D}$ interface across the entire torus. Finally one can contract it again leaving the entire system in $\mathcal{T}_1$ (steps are shown in Fig.~\ref{fig:sweeping}). This establishes the equality of partition functions of  $\mathcal{T}_0$  and  $\mathcal{T}_1$, including in the presence of any background non-contractible $G$ network. 
(It is straightforward to check that all the potential phases from  the $F$-moves of $\cal D$ cancel in this process.)
This implies that for each scaling dimension, the $G$ representation characters are identical, i.e., symmetries act the same way on local and nonlocal operators.

This argument breaks down for non-invertible topological interfaces, for which the necessary $F$-moves generate sums of other defect contributions. 
See Ref.~\onlinecite{Chang:2018iay} for an example involving the non-invertible Kramers-Wannier defect.  
For more generic conformal defects, one cannot topologically deform it   without changing the correlation functions.

We note that as a special case of this argument, we recover the fact that gapped SPT phases protected by internal symmetries must have edge modes. 
In that context, we take $\mathcal T_0$ to be the trivial phase, then the interface can be interpreted as a boundary. 
Indeed, for a gapped phase, the only way that the 0+1d $G$-symmetric interface can be non-invertible is by being degenerate. 
See Appendix \ref{app:gappedSPT}.

%Formally, if $Z[g,h, \lbrace\phi_j(z_j)\rbrace]$ is the partition function with spacelike $g$ and timelike $h$ twist and  $\phi_j(z_j)$ insertions, then 
%\[Z_{0}[g,h, \lbrace\phi_j(z_j)\rbrace] = Z_{0}[g,h, \lbrace \mathcal{D}\phi_j(z_j)\mathcal{D}^{-1}\rbrace] \]

\subsection{Constraints from two-point functions across interfaces between distinct criticalities}
\label{subsec:vanishing_2pt}

Another way to see the defect invertibility relies on the relation between defects and operators. 
(See, for example, Ref.~\onlinecite{Shao:2023gho} for a general discussion.) 
In the IR CFT with relativistic symmetry, one can orient a defect $\cal D$ to extend along the time direction, or along the spatial direction. 
In the latter case, this defines a conserved operator, denoted as $\widehat{\cal D}$, acting on the bulk Hilbert space. 
Under the folding trick, $\widehat{\cal D}$ is  mapped to the Cardy boundary state of the doubled CFT~\cite{Ishibashi_1989,CARDY1989581, CARDY1991274,cardy2008boundaryconformalfieldtheory, Cardy17}. 
If \(\mathcal{D}\) is \(G\)-symmetric, it leads to an operator $\widehat{\mathcal{D}}^{(g)}$ acting on the $g$-twisted Hilbert space (which becomes a twisted boundary state upon folding~\cite{Fukusumi:2021zme,Choi:2024tri}). 
The matrix elements of the operators $\widehat{\cal D}$ and $\widehat{\cal D}^{(g)}$ are related to the two-point functions of (local or non-local) operators straddling across the interface:
\begin{equation}
\begin{split}
\langle \phi_a^\dagger(-x)\phi_b(x)\rangle_{\mathcal{D}} &\propto \frac{\langle \phi_a | \widehat{\mathcal{D}} | \phi_b\rangle}{x^{\Delta_{\phi_a} + \Delta_{\phi_b}}} \\
\langle \mu_a^\dagger(-x)\mu_b(x)\rangle_{\mathcal{D}} &\propto \frac{\langle \mu_a | \widehat{\mathcal{D}}^{(g)} | \mu_b\rangle}{x^{\Delta_{\mu_a} + \Delta_{\mu_b}}}
\end{split}
\label{eq:two-point-functions}
\end{equation}

\begin{figure}
    \centering
\begin{tikzpicture}
  \def\yTop{0.9}
  \def\yBot{-0.8}

  \fill[blue!10] (-4,\yBot) rectangle (0,\yTop);
  \fill[red!10]  (0,\yBot)  rectangle (4,\yTop);

  \draw[very thick, violet] (0,\yBot) -- (0,\yTop);
  \node[violet] at (0.22,0.3) {\Large $\mathcal{D}$};

  \def\yt{0.15}     
  \def\yb{-0.35}    

  \filldraw[black] (-1.7,\yt) circle (2pt);
  \filldraw[black] ( 1.7,\yt) circle (2pt);
  \node[above left]  at (-1.7,\yt) {$\phi^\dagger$};
\node[above right] at ( 1.7,\yt) {$\phi$};

\filldraw[black] (-2,\yb) circle (2pt);
\filldraw[black] ( 2,\yb) circle (2pt);

\node[above left]  at (-2,\yb) {$\mu^\dagger$};
\node[above right] at ( 2,\yb) {$\mu$};

  \draw[decorate, decoration={snake, amplitude=0.45mm, segment length=3.2mm}, thick]
    (-2,\yb) -- (2,\yb);

\end{tikzpicture}
    \caption{\textbf{Selection rules from symmetry across an interface of symmetry-enriched criticalities} 
For a $G$–symmetric interface $\mathcal{D}$, any operator whose $G$–charge differs on the two sides has a vanishing two-point function across the interface. In particular, if the local operator $\phi$ carries different $G$–charges in the two regions, then $\langle \phi^\dagger(-x)\,\phi(x)\rangle=0$ (top row). The same conclusion holds in $G$–twisted sectors: if the twisted primary operator $\mu$ has mismatched $G$-charges across the interface, then $\langle \mu^\dagger(-x)\,\mu(x)\rangle=0$ (bottom row). These vanishing correlators impose constraints on the IR defect.}
    \label{fig:two-point-functions}
\end{figure}
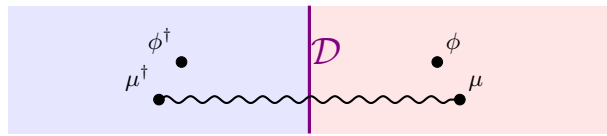

Suppose \(\mathcal{T}_0\) and \(\mathcal{T}_1\) are distinct SECs with some local operators  carrying different \(G\)-charges across the interface. 
It follows that their two-point functions across the interface vanish by symmetry, 
\[ \langle \phi^\dagger(-x) \phi(x)\rangle_{\mathcal{D}} = 0. \]
In the case of degeneracies within a scaling dimension, the corresponding correlation matrix is an intertwiner between the \(G\) representations on either side of the defect on the subspace of that scaling dimension. 
More specifically , let $\rho_0$ and $\rho_1$ be the distinct $G$ representations of the local operators in the two SECs on the two sides. 
Suppose $\widehat{\cal D}$ were invertible, then it would implement an isomorphism between the two representations, $\rho_0 \cong \rho_1$, contradicting the assumption. 
We again arrive at the conclusion that $\cal D$ is non-invertible. 
The same argument applies to the case when the operators under considerations are in the $G$-twisted sector  (see Fig.~\ref{fig:two-point-functions}). In this case, the representations $\rho_0$ and $\rho_1$ are generally projective.

We note that the vanishing of a two-point function, which we inferred in the UV simply from bulk operator charges, is manifested in the IR as an \textit{intrinsic} property of the conformal defect itself, revealed in its universal data. This vanishing imposes strong constraints on the defect. In Sec.~\ref{sec:ising}, we use this constraint to exactly pinpoint the most stable defect in Ising interfaces.

\subsection{Defining symmetry-enriched criticality through non-invertible defects}

Thus far, we have connected the characterization of (symmetry-enriched) criticality in terms of charges of scaling operators to whether or not a spatial interface can be invertible. In the introduction we briefly touched on another viewpoint, namely the definition of symmetry-enriched quantum criticality in terms of paths of local Hamiltonians. The challenge of that existing definition is that it is difficult to prove the (non)existence of such paths. Here we argue that there is a natural conceptual connection between this path-based definition and our new interface-based perspective, suggesting that the latter might be a useful definition that contains the same physical picture but is easier to work with.

Let us recall the path-based definition mentioned in the introduction. Two $G$-symmetric lattice Hamiltonians $H_0$ and $H_1$ flowing to a particular CFT are said to be in the same phase if and only if there exists a $G$-symmetric path of Hamiltonians $H(\lambda)$ with $H(0) = H_0$ and $H(1) = H_1$ such that $H(\lambda)$ flows to the same CFT for every choice of $\lambda$ \cite{Verresen_2021}. If such a family $H(\lambda)$ exists, it seems plausible that one can find a symmetric invertible interface between $H_0$ and $H_1$. A natural candidate is given by tuning through $\lambda(x)$ as a function of space. More precisely, since $H(\lambda)$ flows to the same CFT for each value of $\lambda$, each perturbation must be irrelevant, i.e., $H'(\lambda) \equiv \frac{\rm d}{\rm d \lambda} H(\lambda)$ is an irrelevant operator for $H(\lambda)$. This suggests we can write $H(\lambda) = \sum_x h_x(\lambda)$ such that $h_x' (\lambda)$ is an irrelevant operator. For a choice of $\lambda(x)$ which spatially interpolates from 0 to 1, the Hamiltonian 
\begin{equation}
H_0 + \sum_x \int_0^{\lambda(x)} {\rm d} \mu \; h_x'(\mu) \label{eq:interpol}
\end{equation}
thus defines an interface between $H_0$ (left) and $H_1$ (right). It seems plausible that the RG-irrelevance of $h_x'(\mu)$ can be used to argue that Eq.~\eqref{eq:interpol} flows to $H_0$ for a sufficiently smooth choice of $\lambda(x)$, thereby realizing the identity defect (which, in particular, is invertible); we leave a detailed analysis to future work.

To show the converse direction---namely that the existence of an invertible symmetric interface between $H_0$ and $H_1$ implies the existence of a path $H(\lambda)$---it would be promising to use the ideas of Section~\ref{subsec:sweep}. In particular, the invertible nature of the interface suggests one should be able to nucleate it in pairs, and the topological nature suggests one can sweep it across the lattice, thereby constructing a family of Hamiltonians connecting $H_0$ and $H_1$. However, it is rather challenging to rigorously prove statements about such a path-based definition, which has provided a bottleneck in the study of gapless phases. Our interface-based approach thus provides an attractive alternative, whose definition is likely equivalent to the former, but which is much more manageable to work with, as evidenced by our above arguments, as well as the examples in the next section.

\color{black}

\section{Examples of interfaces between Ising criticalities}
\label{sec:ising}

To make our discussion concrete, we analyze the simplest example of a critical theory—the Ising CFT. The Ising CFT admits several symmetry-enriched realizations on the lattice~\cite{Verresen_2021}. Here, we focus on those protected by a faithful $\mathbb{Z}_2 \times \mathbb{Z}_2^T$ symmetry, generated respectively by the spin-flip operator $\prod_j X_j$ and time-reversal $T$ (complex conjugation in the Pauli-$Z$ basis). Here $X$ and $Z$ refer to the usual Pauli matrices.

All conformal defects of the Ising CFT have been classified in Ref.~\onlinecite{Oshikawa:1996dj}. 
See Appendix \ref{app:Affleck_Oshikawa_defect_classification} for a review. 
Our goal here is to understand which of these defects are actually \emph{compatible with symmetry} when the two sides of the interface correspond to distinct symmetry-enriched realizations.

  A standard representative 
of the Ising universality class
is the transverse-field Ising chain
\begin{equation}
H_{\rm Ising} =  -\sum_j \big(X_j + Z_j Z_{j+1}\big).
\end{equation}
Its spin field $\sigma \sim Z_j$ and disorder field $\mu \sim \prod_{k<j}X_j$ (both with scaling dimension $\Delta = \frac{1}{8}$) are even under time-reversal.  
A closely related symmetry enrichment is obtained with a $\mathbb Z_2^T$-odd spin, i.e., $\sigma \sim Y_j$ \cite{Verresen_2021}:
\begin{equation}
H_{{\rm Ising}}^{\sigma} =  -\sum_j \big(X_j + Y_j Y_{j+1}\big).
\end{equation}
In this case, the two systems do not differ by a gapped SPT; rather, the UV symmetries are embedded into the IR symmetry differently.

A more subtle nonlocal enrichment is the `gapless SPT' studied in Ref.~\onlinecite{Verresen_2021}:
\begin{equation}
H_{\rm Ising}^\mu = -\sum_j \big(Z_j Z_{j+1} + Z_{j-1} X_j Z_{j+1}\big).
\end{equation}
It is related to $H_{\rm Ising}$ by acting the $\mathbb{Z}_2\times \mathbb{Z}_2^T$ entangler $\prod_n \mathrm{CZ}_{n,n+1} = e^{i\frac{\pi}{4} \sum_n (-1)^n Z_n Z_{n+1}}$. 
This model thus shares the same \textit{local} operator charges as $H_{\rm Ising}$, but its Hermitian disorder operator
$\mu \sim \cdots X_{j-3} X_{j-2} Y_{j-1} Z_j$ is odd under time-reversal, i.e., $T \mu T = - \mu$.  
$H_\text{Ising}$ and $H_\text{Ising}^\mu$ differ by a gapped $\mathbb{Z}_2\times \mathbb{Z}_2^T$ SPT, which is described by the continuum 1+1d topological action $i \pi \int w_1 \cup A$. Here $w_1$ is the first Stiefel-Whitney class of the tangent bundle and $A$ is the background $\mathbb{Z}_2$ gauge field.  
In what follows, we will study conformal interfaces between these symmetry-enriched critical points.

\subsection{Interface of $H_{\rm Ising}$ and $H_{\rm Ising}^\sigma$}
\label{subsec:Hs_H}

We begin with the interface between the two simplest realizations, $H_{\rm Ising}$ and $H_{\rm Ising}^{\sigma}$.  
Explicitly, we consider
\begin{equation}
    H = H_{\mathrm{left}}+H_{\mathrm{int}}+H_{\mathrm{right}}
\label{eq:Hs_H}
\end{equation}
with 
\begin{equation}
\begin{split}
    H_{\mathrm{left}} &= \ldots -X_{-2} - Y_{-2} Y_{-1} - X_{-1},\\
H_{\mathrm{int}} &= - b Y_{-1} Y_0 - h X_0 ,\\
H_{\mathrm{right}} &= -Z_0 Z_1 - X_1 - Z_1 Z_2 - X_2 - \cdots   
\end{split} 
\label{eq:Hs_H_detail}
\end{equation}
The simplest choice is $b=0$ and $h \neq 0$, in which case the two halves are decoupled and both flow to the free boundary condition, denoted as $\ket{f}$, of the Ising CFT~\cite{CARDY1989581, CARDY1991274, cardy2008boundaryconformalfieldtheory}. 
This factorized $\ket{f}\bra{f}$ defect is a special case of the continuous family of Dirichlet conformal defects $D(\phi)$ with $\phi=\pi/2$ \cite{Oshikawa:1996dj}. 
It is  locally stable from this microscopic realization: the $\mathbb Z_2 \times \mathbb Z_2^T$ symmetry forbids the defect marginal operator which tunes along the Dirichlet defect moduli space. 
To show this, we denote the $\Delta=1/2$ operator on the free boundary as $\sigma^{(f)}$, which arises from the bulk-boundary OPE with the bulk spin operator $\sigma$. Hence $\sigma^{(f)}$ on either side carries the same $\mathbb{Z}_2\times \mathbb{Z}_2^T$ charges, i.e., $(-,-)$, as its bulk counterpart $\sigma$. 
Since the $\mathbb{Z}_2$-charges of $\sigma$ on the two sides are different, the defect marginal deformation $\sigma^{(f)}_L(0) \sigma^{(f)}_R(0) \sim Y_{-1} Z_0$ is $\mathbb{Z}_2$-odd, and is therefore forbidden. 

Importantly, this implies that we cannot continuously deform $\ket{f}\bra{f}$  to either the identity defect $D(\pi/4)$ or the $\mathbb{Z}_2$ defect $D(3\pi/4)$, which are the two invertible defects  on the Dirichlet moduli space. 
This is consistent with our main statement that two distinct SECs cannot be separated by a symmetric, invertible defect.

However, this is not the only $G$-symmetric defect that can arise at this interface. 
For instance, if $h =b=0$, the defect factorizes into a free boundary $\ket{f}$ on the left, and a   `spontaneously fixed' boundary $\bra{\uparrow}+\bra{\downarrow}$ on the right, labeled by $Z_0=\pm1$.  
This  defect is a special case of the Neumann conformal defect $N(\phi)$ for $\phi=0$ \cite{Oshikawa:1996dj, Bachas_2013, Kormos_2009}, i.e., 
\begin{equation}\label{N0}
N(0)=\ket{f}\bra{\uparrow}+\ket{f}\bra{\downarrow}. 
\end{equation}
We refer to an interface or a defect as non-degenerate (or simple \cite{Choi_2023}) if it has a unique $\Delta=0$ (identity) operator living on it in the IR CFT. 
By the operator-state correspondence, it is equivalent to demanding a non-degenerate ground state in the Hilbert space on a ring with the insertion of the interface and its CPT conjugate. 
In this sense, the interface $N(0)$ is two-fold degenerate. 
The Neumann conformal defects all have a higher Affleck-Ludwig entropy \cite{Affleck_1991}  $g=\sqrt{2}$ (which equals the quantum dimension for topological defects) compared to the Dirichlet defects with $g=1$. 

Turning on $b\neq 0$, the coefficient of $Y_{-1}Y_0$, while keeping $h =0$ corresponds to a defect marginal deformation that is  $\mathbb Z_2 \times \mathbb Z_2^T$-symmetric. 
Denote the  two $\Delta=1/2$ boundary changing operators between the $\bra{\uparrow}$ and $\bra{\downarrow}$ boundaries as $\mu^{\uparrow\downarrow}$ and $\mu^{\downarrow\uparrow}$ \cite{CARDY1989581}, which are exchanged by $\mathbb{Z}_2$. 
Since  $(\mu^{\uparrow\downarrow})^\dagger =\mu^{\downarrow\uparrow}$, the two hermitian operators are $\mu^{\uparrow\downarrow} + \mu^{\downarrow\uparrow}$ and $i(\mu^{\uparrow\downarrow} - \mu^{\downarrow\uparrow})$, which carry $(+,+)$ and $(-,-)$ charges under $\mathbb{Z}_2\times \mathbb{Z}_2^T$, respectively.  The symmetric defect marginal operator is 
\begin{equation}
i\sigma^{(f)}(\mu^{\uparrow\downarrow} - \mu^{\downarrow\uparrow})
\label{eqn:marginaldefectoperator}
\end{equation}
in the continuum. 
Setting $b = \tan \phi$, the microscopic interface realizes the entire branch of  the Neumann defect  $N(\phi)$ of the Ising CFT. 
(See Eq.~(6.10) in Ref.~\onlinecite{Oshikawa:1996dj} after an on-site rotation.)
In particular, when $b=1$,  it becomes a  \textit{non-invertible topological} defect  between $H_{\rm Ising}$ and $H^\sigma_{\rm Ising}$, namely the Kramers-Wannier duality defect $N(\pi/4)$ \cite{Grimm:1992ni,Oshikawa:1996dj,Petkova_2001,Frohlich:2004ef,Aasen:2016dop,Chang:2018iay,Seiberg:2023cdc,Shao:2023gho,Seiberg:2024gek, ueda2025perfectparticletransmissionduality}.

\begin{figure}[t]
\centering
\setlength{\tabcolsep}{2pt}
\renewcommand{\arraystretch}{1.0}

\setlength{\tabcolsep}{3pt}

\setlength{\tabcolsep}{3pt}
\begin{tabular}{cc}

\begin{tikzpicture}[line cap=round, line join=round, scale=0.72, transform shape]
  \def\L{3.0}\def\midW{1.90}\def\midH{0.95}\def\pad{0.0}\def\yoff{0.62}
  \node[anchor=west] at (-\L,  1.5) {\textbf{(a)}};
  \draw[<-, brown!65!black] (-\L,0) -- (-0.5*\midW,0);
  \draw[->, blue!80] ( 0.5*\midW,0) -- ( \L,0);
  \draw[rounded corners=2pt, line width=0.9pt]
    (-0.5*\midW,-0.5*\midH) rectangle (0.5*\midW,0.5*\midH);
  \node at (0, 0.18) {\footnotesize $-\;b\,Y_{-1}Y_{0}$};
  \node at (0,-0.18) {\footnotesize $-h\,X_{0}$};
  \node[anchor=east, brown!65!black] at (-0.5*\midW-\pad,  \yoff) {$-YY-X$};
  \node[anchor=east] at (-0.5*\midW-\pad, -\yoff) {$T\sigma T=-\sigma$};
  \node[anchor=west, blue!80] at ( 0.5*\midW+\pad,  \yoff) {$-ZZ-X$};
  \node[anchor=west] at ( 0.5*\midW+\pad, -\yoff) {$T\sigma T=\sigma$};
\end{tikzpicture}
&

\begin{tikzpicture}[scale=1.0,>=stealth]
 
  \def\xmin{-2.0}\def\xmax{2.0}\def\ymin{-0.4}\def\ymax{1.6}

  \tikzset{plotlab/.style={font=\scriptsize}}
  \colorlet{Nlow}{cyan}
  \colorlet{Nhigh}{blue}
  \colorlet{Afill}{gray!12}

  \begin{scope}
    \clip (\xmin,\ymin) rectangle (\xmax,\ymax);

    \fill[Afill] (\xmin,\ymin) rectangle (0,\ymax);
    \fill[Afill] (0,\ymin) rectangle (\xmax,\ymax);

    \def\w{1.6pt}
    \shade[shading=axis, shading angle=90,
           bottom color=Nlow, top color=Nhigh, draw=none]
          (-0.5*\w, 0) rectangle (0.5*\w, \ymax*0.91);
    \shade[shading=axis, shading angle=90,
           bottom color=Nhigh, top color=Nlow, draw=none]
          (-0.5*\w, \ymin*0.7) rectangle (0.5*\w, 0);
  \end{scope}

  \draw[<->] (\xmin,0) -- (\xmax,0) node[plotlab, above left] {$h$};
  \draw[<->] (0,\ymin) -- (0,\ymax)
  node[pos=0.94, anchor=west, xshift=-1pt, yshift=-1pt, plotlab] {$b=\tan\phi$};

  \node[plotlab, blue!70!black] at (0.35,0.95) {$N(\phi)$};
  \node[plotlab, black!60!black] at (-1.2,0.68) {$ff$};
  \node[plotlab, black!60!black] at ( 1.2,0.68) {$ff$};
 
  \draw[thin] (\xmin,\ymin) rectangle (\xmax,\ymax);
\end{tikzpicture}
\\[6pt]

\begin{tikzpicture}[line cap=round, line join=round, scale=0.72, transform shape]
  \def\L{3.0}\def\midW{1.90}\def\midH{0.95}\def\pad{0.0}\def\yoff{0.62}
\node[anchor=west] at (-\L,  1.5) {\textbf{(b)}};
  \draw[<-, blue!80] (-\L,0) -- (-0.5*\midW,0);
  \draw[->, red!70] ( 0.5*\midW,0) -- ( \L,0);
  \draw[rounded corners=2pt, line width=0.9pt]
    (-0.5*\midW,-0.5*\midH) rectangle (0.5*\midW,0.5*\midH);
  \node at (0, 0.18) {\footnotesize $-\;b\,Z_{-1}X_{0}Z_{1}$};
  \node at (0,-0.18) {\footnotesize $-\;h\,Z_{-1}Z_{0}$};
  \node[anchor=east, blue!80] at (-0.5*\midW-\pad,  \yoff) {$-ZZ-X$};
  \node[anchor=east] at (-0.5*\midW-\pad, -\yoff) {$T\mu T=\mu$};
  \node[anchor=west, red!70] at ( 0.5*\midW+\pad,  \yoff) {$-ZZ-ZXZ$};
  \node[anchor=west] at ( 0.5*\midW+\pad, -\yoff) {$T\mu T=-\mu$};
\end{tikzpicture}
&
% --- Right cell: Plot 2 (wide, shallow; clipped) ---
\begin{tikzpicture}[scale=1.0,>=stealth]
  \def\xmin{-2.0}\def\xmax{2.0}\def\ymin{-0.4}\def\ymax{1.6}
  \tikzset{plotlab/.style={font=\scriptsize}}
  \colorlet{Nlow}{cyan}
  \colorlet{Nhigh}{blue}
  \colorlet{Bfill}{red!15}
  \colorlet{Cfill}{red!35}

  \begin{scope}
    \clip (\xmin,\ymin) rectangle (\xmax,\ymax);
    \fill[Bfill] (\xmin,\ymin) rectangle (0,\ymax);
    \fill[Cfill] (0,\ymin) rectangle (\xmax,\ymax);
    \def\w{1.6pt}
    \shade[shading=axis, shading angle=90,
           bottom color=Nlow, top color=Nhigh, draw=none]
          (-0.5*\w, 0) rectangle (0.5*\w, \ymax*0.91);
    \shade[shading=axis, shading angle=90,
           bottom color=Nhigh, top color=Nlow, draw=none]
          (-0.5*\w, \ymin*0.7) rectangle (0.5*\w, 0);
  \end{scope}

  \draw[<->] (\xmin,0) -- (\xmax,0) node[plotlab, above left] {$h$};
  \draw[<->] (0,\ymin) -- (0,\ymax)
  node[pos=0.94, anchor=west, xshift=-1pt, yshift=-1pt, plotlab] {$b=\tan\phi$};

  \node[plotlab, red!60!black] at (-1.15,0.68) {$\uparrow\downarrow+\downarrow\uparrow$};
  \node[plotlab, red!60!black] at ( 1.15,0.68) {$\uparrow\uparrow+\downarrow\downarrow$};

  \draw[thin] (\xmin,\ymin) rectangle (\xmax,\ymax);
\end{tikzpicture}
\\[6pt]

\begin{tikzpicture}[line cap=round, line join=round, scale=0.72, transform shape]
  \def\L{3.0}\def\midW{1.90}\def\midH{0.95}\def\pad{0.0}\def\yoff{0.62}
  \node[anchor=west] at (-\L,  1.5) {\textbf{(c)}};
  \draw[<-, brown!65!black] (-\L,0) -- (-0.5*\midW,0);
  \draw[->, red!70] ( 0.5*\midW,0) -- ( \L,0);
  \draw[rounded corners=2pt, line width=0.9pt]
    (-0.5*\midW,-0.5*\midH) rectangle (0.5*\midW,0.5*\midH);
  \node at (0, 0.18) {\footnotesize $-\;h_{X}\,X_{0}$};
  \node at (0,-0.18) {\footnotesize $-\;h_{Z}\,Z_{0}Z_{1}$};

  \node[anchor=east, brown!65!black] at (-0.5*\midW-\pad,  \yoff) {$-YY-X$};
  \node[anchor=east] at (-0.5*\midW-\pad, -\yoff) {$T\sigma T=-\sigma$};
  \node[anchor=west, red!70] at ( 0.5*\midW+\pad,  \yoff) {$-ZZ-ZXZ$};
  \node[anchor=west] at ( 0.5*\midW+\pad, -\yoff) {$T\sigma T=+\sigma$};

  \node[anchor=east] at (-0.5*\midW-\pad, -\yoff-0.36) {$T\mu T=+\mu$};
  \node[anchor=west] at ( 0.5*\midW+\pad, -\yoff-0.36) {$T\mu T=-\mu$};
\end{tikzpicture}
&

\begin{tikzpicture}[scale=1.0,>=stealth]
  \def\xmin{-2.0}\def\xmax{2.0}\def\ymin{-1.0}\def\ymax{1.0}
  \tikzset{plotlab/.style={font=\scriptsize}}
  \colorlet{Nmid}{cyan}
  \colorlet{Ddot}{red!60}

  \begin{scope}
    \clip (\xmin,\ymin) rectangle (\xmax,\ymax);
    \fill[Nmid!12] (\xmin,\ymin) rectangle (\xmax,\ymax);
    \filldraw[draw=Ddot!80!black, fill=Ddot, line width=0.6pt] (0,0) circle (2.0pt);
  \end{scope}

\node[plotlab, Ddot!80!black] at (0.45,0.16) 
  {$(\!\uparrow\!+\!\downarrow\!)^{2}$};
  \node[plotlab, blue!70!black] at (-1.15,0.68) {$f\uparrow+f\downarrow$};

  \draw[<->] (\xmin,0) -- (\xmax,0) node[plotlab, above left] {$h_X$};
  \draw[<->] (0,\ymin) -- (0,\ymax)
  node[pos=0.94, anchor=west, xshift=-1pt, yshift=-1pt, plotlab] {$h_Z$};
  \draw[thin] (\xmin,\ymin) rectangle (\xmax,\ymax);
\end{tikzpicture}
\\
\end{tabular}

\caption{
\textbf{Interface phase diagrams for Ising symmetry--enriched criticalities.}
(a) Interface between $H_{\rm Ising}^\sigma$ and $H_{\rm Ising}$ (different $\sigma$ charge across the interface). Tuning $b$ moves along the Neumann family $N(\phi)$ with defect entropy $g = \sqrt{2}$, crossing the Kramers-Wannier defect at $b=1$; any $h \neq 0$ drives a flow to a factorizing defect with free boundaries on both sides ($g=1$).
(b) Interface between $H_{\rm Ising}$ and $H_{\rm Ising}^\mu$ (different $\mu$ charge). 
The spins in the two stable phases  are spontaneously aligned for $h>0$ and anti-aligned for $h<0$.
(c) Interface between $H_{\rm Ising}^\sigma$ and $H_{\rm Ising}^\mu$ (different $\mu$ and $\sigma$ charge). 
The fine-tuned spontaneously fixed $g=2$ defect is unstable and any relevant perturbation flows to a factorizing $N(0)$ interface (free on one side, spontaneously fixed on the other). Here $h_X = h \cos \theta$ and $h_Z = h \sin \theta$.
\label{fig:Ising_interface}}
\end{figure}
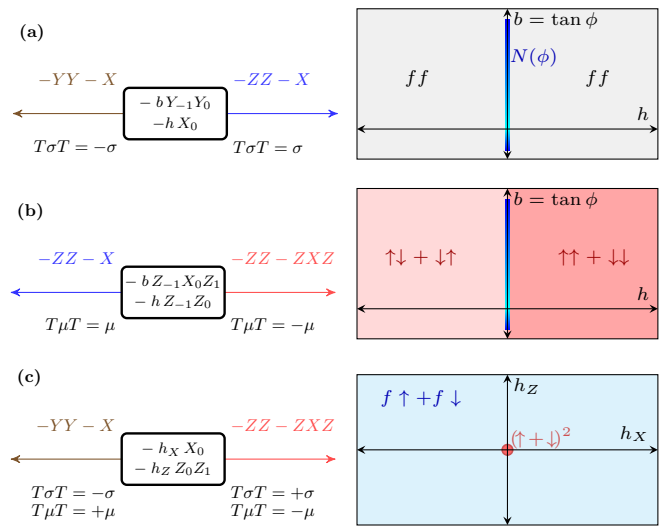

For any $b$, turning on $h\neq 0$, the coefficient of $X_0$, is a relevant defect perturbation. 
In the continuum, it corresponds to the $\mathbb{Z}_2\times \mathbb{Z}_2^T$ symmetric  operator $\mu^{\uparrow\downarrow} +\mu^{\downarrow\uparrow}$ on the free boundary, which arises from the bulk-boundary OPE with the bulk disorder operator $\mu$.\footnote{Denote the $\Delta=0$  operators on the $\bra{\uparrow}$ and $\bra{\downarrow}$ boundaries as $\mathbb{I}^{\uparrow}$ and $\mathbb{I}^{\downarrow}$, respectively. 
Then the lattice operator $Z_0$  flows to  $\mathbb{I}^{\uparrow} - \mathbb{I}^{\downarrow}$, which  is   the order parameter for this spontaneously fixed  boundary $\bra{\uparrow}+\bra{\downarrow}$. }
This deformation drives the Neumann defect $N(\phi)$  back to  $\ket{f}\bra{f}$  with $g=1$. 
We summarize the resulting defect phase diagram in Fig.~\ref{fig:Ising_interface}(a), with a numerical confirmation of the RG flow in Fig.~\ref{fig:N_to_ff}.

More generally, $H_{\mathrm{int}}$ could contain \textit{any} symmetry-allowed terms supported on finitely many sites. 
The two-parameter diagram of Fig.~\ref{fig:Ising_interface}(a) captures all the conformal interface universality classes that can appear without additional fine-tuning. 
In particular, among all the $g=1$ Dirichlet defects~\cite{Oshikawa:1996dj}, only $\ket{f}\bra{f}$ can appear since it is the only one with a vanishing two-point function  $\langle \sigma(-x)\sigma(x)\rangle=0$,\footnote{The correlation function $\langle \sigma(-x)\sigma(x)\rangle$ across Dirichlet defects $D(\phi)$ is found explicitly in Ref.~\onlinecite{Oshikawa:1996dj}; the $\phi$ dependence enters through a an elliptic theta function factor $\vartheta_2(e^{2i\phi},u)$ where $u$ parametrizes $x$. This factor vanishes if and only if $e^{2i\phi}=-1$ corresponding to the $\ket{f}\bra{f}$ defect.} which is required by time-reversal symmetry (see Sec. \ref{subsec:vanishing_2pt}). 
In Sec. \ref{subsec:projective_ising} we present an alternative proof of this based on the defect's endpoint Hilbert space.

\subsection{Interface of $H_{\rm Ising}$ and $H^{\mu}_{\rm Ising}$}
\label{subsec:H_Hmu}

We now turn to the interface between $H_{\rm Ising}$ and the  nonlocally enriched $H^{\mu}_{\rm Ising}$,
\begin{align*}
H \;=\; H_{\mathrm{left}} + H_{\mathrm{int}} + H_{\mathrm{right}},
\end{align*}
with
\begin{equation}
\begin{split}
H_{\mathrm{left}}  &= \cdots - Z_{-3} Z_{-2} - X_{-2} - Z_{-2} Z_{-1} - X_{-1},\\
H_{\mathrm{int}}   &= -\,b\, Z_{-1} X_0 Z_1 - h\, Z_{-1} Z_0, \\
H_{\mathrm{right}} &= \;\; -Z_0 Z_1 - Z_0 X_1 Z_2 - Z_1 Z_2 - \cdots
\end{split}
\label{eq:H_Hu_detail}
\end{equation}

This setup is related to the interface between 
 $H_\text{Ising}$ and $H_\text{Ising}^\sigma$ by  Kramers-Wannier transformation (together with an exchange of the left and right sides), which corresponds to gauging the $\mathbb{Z}_2$ symmetry generated by $\prod_j X_j$. 
 This permutes the charge and twisted sectors. 
 At $h=0$ and $b=\tan\phi$, it realizes the Neumann family $N(\phi)$. 
 From any $N(\phi)$, turning on $h>0$ ($h<0$) drives a relevant flow to the stable, spontaneously aligned (anti-aligned) Dirichlet defect $D(0)$ ($D(\pi)$). 
Both of these two defects are degenerate and factorize into a pair of boundary conditions, i.e.,
\begin{equation}
D(0) =  \ket{\uparrow} \bra{\uparrow}  +  \ket{\downarrow}\bra{\downarrow}  ~~,~~D(\pi) =  \ket{\uparrow}\bra{\downarrow} +  \ket{\downarrow}\bra{\uparrow}.
\end{equation}
Each of these interfaces spontaneously picks one of two spin configurations. 
Intriguingly, there is a direct continuous boundary phase transition between $D(0)$ and $D(\pi)$ at the Neumann line ($h=0$).
In terms of degeneracy on a finite ring (with two interfaces), the $h \neq 0$ case has a two-fold degeneracy, which gets lifted at $h=0$ for generic $b$.\footnote{Even though the degeneracy with two interfaces is lifted, this family of interfaces still has a defect 't Hooft anomaly \cite{Antinucci:2024izg,Komargodski:2025jbu} across the transition, which will be discussed in Sec.~\ref{sec:projective}.}
This can be
interpreted as a 0+1d transition between two distinct symmetry-breaking patterns, akin to the 0+1d ‘DQCP’ studied in Ref.~\onlinecite{Prembabu_2024}.

\begin{figure}[t]
    \centering

    \begin{minipage}[b]{0.48\linewidth}
    \centering
    \begin{tikzpicture}[line cap=round,line join=round,scale=0.45,transform shape]
    \colorlet{sptzero}{blue!80}   
    \colorlet{sptone}{red!70}     
    \tikzset{
      Tlab/.style={font=\Large},
      DefLab/.style={font=\Large,text=orange!85!black},
      GLab/.style={font=\Large},
      SOp/.style={font=\Large},
    }
    \def\ax{3.2}
    \def\ay{1.9}
    \def\TopY{1.06}
    \def\BotY{-1.06}
\coordinate (CL) at (0,0);

    \draw[line width=1pt,sptzero] (CL) ++(0:\ax)   arc[start angle=0,end angle=180,x radius=\ax,y radius=\ay];
    \draw[line width=1pt,sptone]  (CL) ++(180:\ax) arc[start angle=180,end angle=360,x radius=\ax,y radius=\ay];

    \node[Tlab,sptzero] at ($(CL)+(0,\TopY)$) {$H$};
    \node[Tlab,sptone]  at ($(CL)+(0,\BotY)$) {$H^\mu$};

    \path (CL) ++(180:\ax) coordinate (LhL); 
    \path (CL) ++(  0:\ax) coordinate (LhR); 

    \begin{scope}[shift={(LhL)}, rotate=20]
      \draw[->,line width=1.0pt,sptzero] (-0.05, 0.22) -- (0.95, 0.22);
      \draw[->,line width=1.0pt,sptone ] (-0.05,-0.22) -- (0.95,-0.22);
    \end{scope}
   
    \begin{scope}[shift={(LhR)}, rotate=20]
      \draw[->,line width=1.0pt,sptzero] (-0.05, 0.22) -- (0.95, 0.22);
      \draw[->,line width=1.0pt,sptone ] (-0.05,-0.22) -- (0.95,-0.22);
    \end{scope}
    \end{tikzpicture}
    \begin{tikzpicture}[overlay, remember picture]
        \node[font=\bfseries, anchor=west] at (-3.6,1.8) {(a)};
    \end{tikzpicture}
    \end{minipage}\hfill
    \begin{minipage}[b]{0.48\linewidth}
    \centering
    \begin{tikzpicture}[line cap=round,line join=round,scale=0.45,transform shape]
    \colorlet{sptsigma}{brown!65!black}     
    \colorlet{sptone}{red!70}     
    \tikzset{
      Tlab/.style={font=\Large},
      DefLab/.style={font=\Large,text=orange!85!black},
      GLab/.style={font=\Large},
      SOp/.style={font=\Large},
    }
   
    \def\ax{3.2}
    \def\ay{1.9}
    \def\TopY{1.06}
    \def\BotY{-1.06}

    \coordinate (CR) at (0,0);

    \draw[line width=1pt,sptsigma] (CR) ++(0:\ax)   arc[start angle=0,end angle=180,x radius=\ax,y radius=\ay];   
    \draw[line width=1pt,sptone]   (CR) ++(180:\ax) arc[start angle=180,end angle=360,x radius=\ax,y radius=\ay]; 
    \node[brown!65!black] at (-2.8,0.3) {\Large $f$};
    \node[brown!65!black] at (2.8,0.3) {\Large $f$};

    \node[Tlab,sptsigma] at ($(CR)+(0,\TopY)$) {$H^\sigma$}; 
    \node[Tlab,sptone]   at ($(CR)+(0,\BotY)$) {$H^\mu$};

    \path (CR) ++(180:\ax) coordinate (RhL); 
    \path (CR) ++(  0:\ax) coordinate (RhR);

    \begin{scope}[shift={(RhL)}, rotate=20]
      \fill[sptsigma] (0.0, 0.0) circle[radius=5pt];         
      \draw[->,line width=1.0pt,sptone] (-0.05,-0.22) -- (0.95,-0.22); 
    \end{scope}
  
    \begin{scope}[shift={(RhR)}, rotate=20]
      \fill[sptsigma] (0.0, 0.0) circle[radius=5pt];         
      \draw[->,line width=1.0pt,sptone] (-0.05,-0.22) -- (0.95,-0.22); 
    \end{scope}
    \end{tikzpicture}
    \begin{tikzpicture}[overlay, remember picture]
        \node[font=\bfseries, anchor=west] at (-3.6,1.8) {(b)};
    \end{tikzpicture}
    \end{minipage}

    \vspace{0.6em}

    \begin{minipage}[b]{0.48\linewidth}
      \centering
      \includegraphics[width=\linewidth]{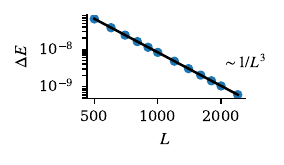}
    \end{minipage}\hfill
    \begin{minipage}[b]{0.48\linewidth}
      \centering
      \includegraphics[width=\linewidth]{gap_L3.pdf}
    \end{minipage}

   \caption{\textbf{Algebraically localized stable interface modes.}
Interfaces between certain symmetry-enriched criticalities can host stable, degenerate modes with spontaneous symmetry breaking (SSB). Shown for Ising criticalities enriched with $\mathbb{Z}_2 \times \mathbb{Z}_2^T$ symmetry, where twisted-sector operators carry different $\mathbb{Z}_2$-charges across the interface. Localization is quantified by the scaling of finite-size ground-state splitting $\Delta E(L)$ on a periodic chain of length $L$, computed by exact diagonalization using a free-fermion mapping.
\textbf{(a)} $H_{\rm Ising}$ and $H_{\rm Ising}^{\mu}$ realize a stable, degenerate interface with  $\Delta E \sim L^{-3}$ (computed for  $h=b=1$ from Eq.~\eqref{eq:H_Hu_detail}).
\textbf{(b)} $H_{\rm Ising}^\sigma$ and $H_{\rm Ising}^{\mu}$ \textit{always} realize a degenerate conformal interface; in the most stable case the gap also scales as $\Delta E \sim L^{-3}$ (computed here for $h = 1$, $\theta=0$ from Eq.~\eqref{eq:Hs_Hmu_detail} with additional irrelevant lattice perturbation $-Y_0 Y_1$). 
\label{fig:gap_L3_L6}}
\end{figure}

On a ring of length $L$ with two $D(0)$ interfaces, the spontaneously-aligned ground-state energy splitting scales as $L^{-3}$, indicating \textit{algebraic} localization (Fig.~\ref{fig:gap_L3_L6}(a)). 
This scaling reflects the underlying spectrum of operators living at the defect site, which can by deduced from the boundary CFT formalism~\cite{Verresen_2021}.  
A straightforward generalization of the analysis in Ref.~\onlinecite{CARDY1989581} shows that there are two $\Delta=1$ defect-changing-operators on $D(0)$, which are $O=\mu^{\uparrow\downarrow}_L \mu^{\uparrow\downarrow}_R$   and $O^\dagger=\mu^{\downarrow\uparrow}_L \mu^{\downarrow\uparrow}_R$.  
Since the disorder operators on the two sides carry opposite $\mathbb{Z}_2^T$-charges, the combination $O+O^\dagger$ (corresponding to $Y_0Z_1$) is $\mathbb{Z}_2$-even but $\mathbb{Z}_2^T$-odd. 
On the other hand, the hermitian operator $i( O - O^\dagger)$ (corresponding to $X_0Z_1$) is $\mathbb{Z}_2^T$-even but $\mathbb{Z}_2$-odd. 
The lightest $\mathbb{Z}_2\times\mathbb{Z}_2^T$-symmetric Hermitian operator is a level-1 descendant, which takes the form (modulo redundant operators which are total derivatives)
\[
\mu^{\uparrow\downarrow}_L \partial_t \mu^{\uparrow\downarrow}_R
+\mu^{\downarrow\uparrow}_L \partial_t \mu^{\downarrow\uparrow}_R ,
\]
with $\Delta=2$. This explains the  $\sim 1/L^{2\Delta-1} = 1/L^3$ finite-size splitting \cite{Verresen_2021}.
A similar analysis applies to $D(\pi)$.

\subsection{Interface of $H^{\sigma}_{\rm Ising}$ and $H^\mu_{\rm Ising}$}
\label{subsec:Hs_Hmu_interface}

The interface between $H^{\sigma}_{\rm Ising}$ and $H^\mu_{\rm Ising}$ is more constrained, with both the local $\sigma$ and nonlocal $\mu$ operators having different symmetry charges across the interface. A convenient lattice realization is 
\begin{equation}
\begin{aligned}
H_{\mathrm{left}}  &= \cdots \;-\; X_{-2} \;-\; Y_{-2}Y_{-1} \;-\; X_{-1} \;-\; Y_{-1}Y_{0} , \\
H_{\mathrm{int}}   &= -h(\cos \theta \, X_0+ \sin\theta\, Z_0 Z_1) , \\
H_{\mathrm{right}} &= -Z_1 Z_2 \;-\; Z_1 X_2 Z_3 \;-\; Z_2 Z_3 \;-\; \cdots
\label{eq:Hs_Hmu_detail}
\end{aligned}
\end{equation}

At $h=0$ the interface is spontaneously fixed on both sides. Turning on $h$ (for any $\theta$) drives an RG flow to the degenerate, factorizing Neumann defect $N(0)=\ket{f} \bra{\uparrow} + \ket{f}\bra{\downarrow}$ in Eq. \ref{N0} with $g=\sqrt{2}$  
(Fig.~\ref{fig:Ising_interface}(c)). To see this, first observe that by conjugating with $e^{i \frac{\theta}{2} Y_0 Z_1}$, we can restrict to $\theta=0$. The two chains then decouple and we readily recognize their boundary conditions.

Symmetry forbids any relevant perturbation of the $N(0) $ interface, as well as any marginal deformations to other $g=\sqrt{2}$ defects.\footnote{Note that $T$ acts with on the $\mu$'s with an extra minus sign compared to the interface in Sec. \ref{subsec:Hs_H}. Hence, the $\mathbb{Z}_2\times \mathbb{Z}_2^T$ charges of the relevant operators $\sigma^{(f)}$, $\mu^{\uparrow\downarrow} +\mu^{\downarrow\uparrow}$, $i(\mu^{\uparrow\downarrow} -\mu^{\downarrow\uparrow})$ are $(-,-) , (+,-) , (-,+)$, respectively.  
It follows that the marginal operators $\sigma^{(f)} (\mu^{\uparrow\downarrow} +\mu^{\downarrow\uparrow})$ and 
$i\sigma^{(f)} (\mu^{\uparrow\downarrow} -\mu^{\downarrow\uparrow})$ are also forbidden by the symmetry. }
Thus it hosts robust symmetry-protected degenerate modes in the thermodynamic limit. 
Furthermore, there is \emph{no} symmetry-allowed modification of the defect region that can remove this degeneracy or lower its $g$ function, even non-perturbatively; see Appendix~\ref{app:Hsimga_Hmu_defects_proof}.

On a ring of length $L$ with two interfaces $N(0)$ (or its generic, symmetric deformation), the ground state energy also splits $\propto 1/L^3$.
This can be seen using similar reasoning as in Sec.~\ref{subsec:H_Hmu}. Since the $Y_0 \sim \sigma^{(f)}$ field on the free boundary is odd under both $\mathbb{Z}_2$ and $\mathbb{Z}_2^T$, the lightest Hermitian $\mathbb{Z}_2\times \mathbb{Z}_2^T$-symmetric field coupling the  $\ket{f} \bra{\uparrow}$ and $\ket{f}\bra{\downarrow}$ defects is a Virasoro descendant:
\begin{equation}
i\sigma^{(f)} \partial_t(\mu^{\downarrow \uparrow}-\mu^{\uparrow \downarrow}) 
\end{equation}
It has dimension $\Delta=2$ and as a perturbation leads to a degeneracy splitting $\propto 1/L^{2\Delta-1}$.
Fig.~\ref{fig:gap_L3_L6}(b) shows this scaling 
with the corresponding perturbation term $- Y_0 Y_1$ at the interface site.
%}

As an aside, we explore how the same conformal boundary condition can have sharper algebraic localization with stronger symmetry constraints.
A similar Ising interface setup can be considered with global $\mathbb{Z}_2 \times \mathbb{Z}_2 \times \mathbb{Z}_2^T$ symmetry where the unitary symmetry acts non-faithfully on the Ising CFT, with different sides of the interface transforming under a different $\mathbb{Z}_2$ subgroup. 
For this interface setup, the same $N(0)$ interface is allowed but the finite size splitting is $\propto 1/L^6$.
The larger exponent reflects how only among the level-3 descendants of $\mathbb{Z}_2^T$-odd primary $i(\mu^{\uparrow \downarrow} + \mu^{\downarrow \uparrow})$ do we find the lightest non-redundant $\mathbb{Z}_2 \times \mathbb{Z}_2 \times \mathbb{Z}_2^T$ symmetric perturbation that couples the degenerate ground states.

\subsection{Fermionic symmetry-enriched criticalities}

By performing a Jordan-Wigner (JW) transformation on the above Ising spin chains, we can furnish fermionic examples of interfaces between distinct symmetric quantum chains. In this case, the bulk criticality is that of the Majorana CFT with central charge $c=\frac{1}{2}$.

Globally on a ring, JW maps a local spin chain Hamiltonian to a nonlocal fermionic Hamiltonian. 
To fermionize the bosonic spin chain model globally, one needs to gauge the $\mathbb{Z}_2$ symmetry by introducing certain fermionic degrees of freedom to arrive at a local fermionic model. 
See Appendix B of Ref. \onlinecite{Pace:2024oys} and earlier works in Refs. \onlinecite{Alvarez-Gaume:1986nqf,Alvarez-Gaume:1987wwg,Kapustin:2017jrc,Karch:2019lnn,Ji:2019ugf,Hsieh:2020uwb} on bosonization/fermionization. 
The gauging changes the physics of these interfaces. Indeed, we will see that it  affects the degeneracy of the spectrum.

Let $\{\gamma_j\}_j$ be Majorana fermions with $\{\gamma_i,\gamma_j\}=2\delta_{ij}$. The symmetries are generated by fermion parity $(-1)^F$ and an antiunitary time-reversal $T$ which acts as follows:
\begin{equation}
(-1)^F\gamma_j (-1)^F = -\gamma_j,  \qquad T\gamma_j T^{-1}=(-1)^j\gamma_j .
\end{equation} 
We consider the following models, representing distinct symmetry-enriched realizations of the Majorana CFT~\cite{Verresen_2018,Verresen_2021}:
\begin{equation}
\begin{split}
H_F &= i\sum_j(\gamma_{2j-1}\gamma_{2j}+\gamma_{2j}\gamma_{2j+1}),\\
H_F^{\sigma} &= i\sum_j(\gamma_{2j}\gamma_{2j+3}+\gamma_{2j}\gamma_{2j+1}),\\
H_F^{\mu} &= i\sum_j(\gamma_{2j-1}\gamma_{2j}+\gamma_{2j-3}\gamma_{2j}).\\
\end{split}
\label{eq:fermionic}
\end{equation}
For one particular choice of JW, these map to $H_{\rm Ising}$, $H_{\rm Ising}^\sigma$ and $H_{\rm Ising}^\mu$, respectively. Note $H^\mu_F$ is a spatial reflection of $H^\sigma_F$.

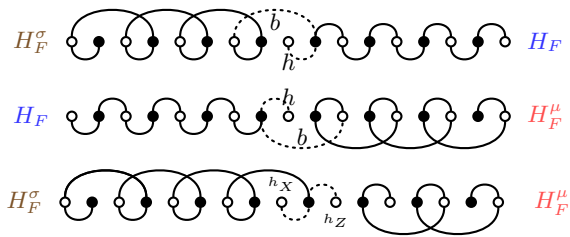
\begin{figure}[t]
\centering

    \colorlet{sptzero}{blue!80}        
    \colorlet{sptmu}{red!70}         
    \colorlet{sptsigma}{brown!65!black}

\begin{tikzpicture}[x=3.6mm, y=3.6mm, line cap=round, line join=round, >=stealth]

  \node[anchor=east, font=\small] at (-7.0,0) {\color{sptsigma} $H_F^{\sigma}$};
  \node[anchor=west, font=\small] at (10.0,0) {\color{sptzero}  $H_F$};

  \begin{scope}[shift={(1.5,0)}]
    \def\hshort{0.8}
    \def\hlong{1.5}
    \foreach \k in {-8,...,8} {
      \ifnum\k>-9
        \pgfmathtruncatemacro{\col}{mod(abs(\k),2)}
        \ifnum\col=0
          \node[circle,fill=white,draw=black,thick,inner sep=1.25pt] (n\k) at (\k,0) {};
        \else
          \node[circle,fill=black,draw=black,thick,inner sep=1.25pt] (n\k) at (\k,0) {};
        \fi
      \else
        \coordinate (n\k) at (\k,0);
      \fi
    }
    \draw[thick] (n-7) .. controls +(0,-\hshort) and +(0,-\hshort) .. (n-8);
    \draw[thick] (n-8) .. controls +(0,\hlong)   and +(0,\hlong)   .. (n-5);
    \draw[thick] (n-5) .. controls +(0,-\hshort) and +(0,-\hshort) .. (n-6);
    \draw[thick] (n-6) .. controls +(0,\hlong)   and +(0,\hlong)   .. (n-3);
    \draw[thick] (n-4) .. controls +(0,-\hshort) and +(0,-\hshort) .. (n-3);
    \draw[thick] (n-4) .. controls +(0,\hlong)   and +(0,\hlong)   .. (n-1);
    \draw[thick] (n-2) .. controls +(0,-\hshort) and +(0,-\hshort) .. (n-1);

    \draw[thick,dotted] (n-2) .. controls +(0,\hlong) and +(0,\hlong) .. (n1);
    \node[font=\small] at ($(n0)-(0.5,-0.75)$) {$b$};

    \draw[thick,dotted] (n0) .. controls +(0,-\hshort) and +(0,-\hshort) .. (n1);
    \node[font=\small, below = 1pt] at (n0) {$h$};

    \draw[thick] (n1) .. controls +(0,\hshort)   and +(0,\hshort)   .. (n2);
    \draw[thick] (n2) .. controls +(0,-\hshort)  and +(0,-\hshort)  .. (n3);
    \draw[thick] (n3) .. controls +(0,\hshort)   and +(0,\hshort)   .. (n4);
    \draw[thick] (n4) .. controls +(0,-\hshort)  and +(0,-\hshort)  .. (n5);
    \draw[thick] (n5) .. controls +(0,\hshort)   and +(0,\hshort)   .. (n6);
    \draw[thick] (n6) .. controls +(0,-\hshort)  and +(0,-\hshort)  .. (n7);
    \draw[thick] (n7) .. controls +(0,\hshort)  and +(0,\hshort)  .. (n8);
  \end{scope}
\end{tikzpicture}
% ---------------- PANEL 2 ----------------
\begin{tikzpicture}[x=3.6mm, y=3.6mm, line cap=round, line join=round, >=stealth]
  \node[anchor=east, font=\small] at (-7.0,0) {\color{sptzero} $H_F$};
  \node[anchor=west, font=\small] at (10.0,0) {\color{sptmu} $H_F^{\mu}$};

  \begin{scope}[shift={(1.5,0)}]
    \def\hshort{0.8}
    \def\hlong{1.5}
    \foreach \k in {-8,...,8} {
      \ifnum\k>-9
        \pgfmathtruncatemacro{\col}{mod(abs(\k),2)}
        \ifnum\col=0
          \node[circle,fill=white,draw=black,thick,inner sep=1.25pt] (n\k) at (\k,0) {};
        \else
          \node[circle,fill=black,draw=black,thick,inner sep=1.25pt] (n\k) at (\k,0) {};
        \fi
      \else
        \coordinate (n\k) at (\k,0);
      \fi
    }
    \draw[thick] (n-8) .. controls +(0,-\hshort)   and +(0,-\hshort)   .. (n-7);
    \draw[thick] (n-7) .. controls +(0,\hshort)    and +(0,\hshort)    .. (n-6);
    \draw[thick] (n-6) .. controls +(0,-\hshort)   and +(0,-\hshort)   .. (n-5);
    \draw[thick] (n-5) .. controls +(0,\hshort)    and +(0,\hshort)    .. (n-4);
    \draw[thick] (n-4) .. controls +(0,-\hshort)   and +(0,-\hshort)   .. (n-3);
    \draw[thick] (n-3) .. controls +(0,\hshort)    and +(0,\hshort)    .. (n-2);
    \draw[thick] (n-2) .. controls +(0,-\hshort)   and +(0,-\hshort)   .. (n-1);

    \draw[thick,dotted] (n-1) .. controls +(0,\hshort) and +(0,\hshort) .. (n0);
    \node[font=\small, above=1pt] at (n0) {$h$};

    \draw[thick,dotted] (n-1) .. controls +(0,-\hlong) and +(0,-\hlong) .. (n2);
    \node[font=\small] at ($(n0)+(0.5,-0.75)$) {$b$};

    \draw[thick] (n1) .. controls +(0,\hshort)   and +(0,\hshort)   .. (n2);
    \draw[thick] (n1) .. controls +(0,-\hlong)   and +(0,-\hlong)   .. (n4);
    \draw[thick] (n3) .. controls +(0,\hshort)   and +(0,\hshort)   .. (n4);
    \draw[thick] (n3) .. controls +(0,-\hlong)   and +(0,-\hlong)   .. (n6);
    \draw[thick] (n5) .. controls +(0,\hshort)   and +(0,\hshort)   .. (n6);
    \draw[thick] (n5) .. controls +(0,-\hlong)   and +(0,-\hlong)   .. (n8);
    \draw[thick] (n7) .. controls +(0,\hshort)   and +(0,\hshort)   .. (n8);

  \end{scope}
\end{tikzpicture}

\begin{tikzpicture}[x=3.6mm, y=3.6mm, line cap=round, line join=round, >=stealth]
  \node[anchor=east, font=\small] at (-7.0,0) {
  \color{sptsigma} $H_F^{\sigma}$};
  \node[anchor=west, font=\small] at (10.5,0) {\color{sptmu} $H_F^{\mu}$};

  \begin{scope}[shift={(1.5,0)}]
    \def\hshort{0.8}
    \def\hlong{1.5}
    \foreach \k in {-8,...,8} {
      \ifnum\k>-9
        \pgfmathtruncatemacro{\col}{mod(abs(\k),2)}
        \ifnum\col=0
          \node[circle,fill=white,draw=black,thick,inner sep=1.25pt] (n\k) at (\k,0) {};
        \else
          \node[circle,fill=black,draw=black,thick,inner sep=1.25pt] (n\k) at (\k,0) {};
        \fi
      \else
        \coordinate (n\k) at (\k,0);
      \fi
    }
    \draw[thick] (n-7) .. controls +(0,-\hshort) and +(0,-\hshort) .. (n-8);
    \draw[thick] (n-8) .. controls +(0,\hlong)   and +(0,\hlong)   .. (n-5);
    \draw[thick] (n-5) .. controls +(0,\hlong)   and +(0,\hlong)   .. (n-8);
    \draw[thick] (n-5) .. controls +(0,-\hshort) and +(0,-\hshort) .. (n-6);
    \draw[thick] (n-6) .. controls +(0,\hlong)   and +(0,\hlong)   .. (n-3);
    \draw[thick] (n-4) .. controls +(0,-\hshort) and +(0,-\hshort) .. (n-3);
    \draw[thick] (n-4) .. controls +(0,\hlong)   and +(0,\hlong)   .. (n-1);
    \draw[thick] (n-2) .. controls +(0,-\hshort) and +(0,-\hshort) .. (n-1);
    \draw[thick] (n-2) .. controls +(0,\hlong)   and +(0,\hlong)   .. (n1);

    \draw[thick,dotted] (n0) .. controls +(0,-\hshort) and +(0,-\hshort) .. (n1);
    \node[font=\tiny, above=1pt] at (n0){$h_X$};

    \draw[thick,dotted] (n1) .. controls +(0,\hshort) and +(0,\hshort) .. (n2);
    \node[font=\tiny, below=1pt] at (n2) {$h_Z$};

    \draw[thick] (n3) .. controls +(0,\hshort)   and +(0,\hshort)   .. (n4);
    \draw[thick] (n3) .. controls +(0,-\hlong)   and +(0,-\hlong)   .. (n6);
    \draw[thick] (n5) .. controls +(0,\hshort)   and +(0,\hshort)   .. (n6);
    \draw[thick] (n5) .. controls +(0,-\hlong)   and +(0,-\hlong)   .. (n8);
    \draw[thick] (n7) .. controls +(0,\hshort)   and +(0,\hshort)   .. (n8);

  \end{scope}
\end{tikzpicture}

\caption{\textbf{Free fermion depictions of the interfaces between symmetry-enriched Majorana CFTs.} Each row shows a schematic of the critical Majorana chain, where a curved edge from node $j$ to $k$ ($j<k$) represents the Majorana bilinear $i\gamma_j \gamma_k$. Each row illustrates the Hamiltonians for distinct free-fermion critical chains and their interfaces. Interfaces between inequivalent chains produce either decoupled chains or unpaired Majorana modes, reflecting factorizing defects or robust nonlocal degeneracies, respectively. The patterns in this free-fermion picture capture the essential defect universality classes that persist even with interactions.}
\label{fig:free-fermions}
\end{figure}

The free-fermion regime is a useful case study for the spatial interfaces between these criticalities. One can view each $\mathbb Z_2^T$-symmetric quadratic Hamiltonian as a bipartite graph of even and odd Majorana nodes (Fig.~\ref{fig:free-fermions}).

Up to symmetric orthogonal transformations, the most stable configuration of the free-fermion interface between $H_F$ and $H_F^{\sigma}$ is a non-degenerate, factorizing defect with $g=1$. 
This defect, which arises from the $\ket{f}\bra{f}$ interface in Sec. \ref{subsec:Hs_H} before JW, is non-invertible, in agreement with our general criterion.

When $h=0$, this interface has  a localized, decoupled Majorana zero mode at an otherwise connected chain  (see  Fig.~\ref{fig:free-fermions}).   
This Majorana zero mode  contributes a factor $g=\sqrt{2}$ and produces a two-fold ground-state degeneracy when we have a pair of them on a ring.\footnote{Note that its bosonic counterpart, the Neumann defect $N(\phi)$ in Sec. \ref{subsec:Hs_H}, is non-degenerate, i.e., the ground state is non-degenerate if we have $N(\phi)$ and its dual on a ring. }  
The special point $h=0$ and $b=1$ (i.e., $\phi=\pi/4$) deserves more discussions. 
At this point, skipping the decoupled Majorana fermion, the interface Hamiltonian is invariant under translation by one Majorana site. 
This can be viewed as a translation symmetry defect, which becomes  an invertible  chiral fermion parity defect  \cite{Seiberg:2023cdc,Seiberg:2025zqx} in the continuum, with a decoupled Majorana zero mode. 
Hence, the composite defect is again non-invertible.  
Indeed, it is known that, under fermionization,  the non-invertible Kramers-Wannier topological defect $N(\pi/4)$ becomes an invertible chiral fermion parity defect \cite{Ji:2019ugf,Lin:2019hks,Kaidi:2021xfk,Shao:2023gho}, plus a decoupled Majorana zero mode. 
A similar analysis applies to the interfaces between $H_F$ and $H_F^\mu$.

Finally, we discuss the free-fermion interfaces between $H_F^\mu$ and $H_F^\sigma$. 
From Fig.~\ref{fig:free-fermions}, we see that these interfaces  always carry at least one unpaired Majorana zero mode, leading to a two-fold degeneracy on a ring when we have a pair of these interfaces. 
Again, the most stable interface is  factorizing and hence non-invertible.
These free-fermion results can be mapped to we found in the interacting bosonic case (after the appropriate nonlocal mapping of charge and twist sectors \cite{Hsieh:2020uwb}) and therefore persist upon including local, symmetry-preserving multi-fermion interactions.

\section{Defect anomalies for gapless SPT interfaces} \label{sec:projective}

\begin{figure}[t]
\centering
\begin{tikzpicture}
\node at (0,0){
\begin{tikzpicture}[line cap=round,line join=round,scale=0.45,transform shape]

\colorlet{sptzero}{blue!80}
\colorlet{sptone}{red!70}
\colorlet{defc}{purple!70!black}

\tikzset{
  Tlab/.style={font=\Large},
  DefLab/.style={font=\Large,text=defc},
  GLab/.style={font=\Large},
  SOp/.style={font=\Large},
}

% ---------- geometry (same eccentricity; outer is scaled copy) ----------
\def\ax{3.2}     % inner (Hamiltonian) x-radius
\def\ay{1.9}     % inner (Hamiltonian) y-radius
\def\scl{1.28}   % outer/inner scale (Symmetry)
\pgfmathsetmacro\bx{\scl*\ax} % outer x-radius
\pgfmathsetmacro\by{\scl*\ay} % outer y-radius

\def\TopY{1.06}          % top label inside inner ellipse
\def\BotY{-1.06}         % bottom label inside inner ellipse
\pgfmathsetmacro\Uy{-\ay - 0.30} % U_g label in the annulus

% ================== LEFT ==================
\coordinate (CL) at (3.0,0);

% symmetry (dashed), same center, slightly larger
\draw[dashed,line width=1pt] (CL) ellipse [x radius=\bx, y radius=\by];

% Hamiltonian (solid): split blue/red
\draw[line width=1pt,sptzero] (CL) ++(0:\ax)   arc[start angle=0,end angle=180,x radius=\ax,y radius=\ay];
\draw[line width=1pt,sptone]  (CL) ++(180:\ax) arc[start angle=180,end angle=360,x radius=\ax,y radius=\ay];

% labels inside inner ellipse
\node[Tlab,sptzero] at ($(CL)+(0,\TopY)$) {$\mathcal{T}_0$};
\node[Tlab,sptone]  at ($(CL)+(0,\BotY)$) {$\mathcal{T}_1$};

% symmetry operator label in annulus (between dashed and solid)
\node[SOp] at ($(CL)+(0,\Uy)$) {$g$};

% endpoint blocks on Hamiltonian ONLY (both labeled D)
\path (CL) ++(180:\ax) coordinate (LhL);
\path (CL) ++(  0:\ax) coordinate (LhR);

\draw[fill=gray!10,rounded corners=2pt]
  ($(LhL)+(-0.24,-0.44)$) rectangle ($(LhL)+(0.44,0.44)$);
\draw[fill=gray!10,rounded corners=2pt]
  ($(LhR)+(-0.44,-0.44)$) rectangle ($(LhR)+(0.24,0.44)$);

% labels INSIDE the left/right rectangles (left panel)
\node[DefLab] at ($(LhL)+(0.10,0)$) {$\mathcal{D}$};   % center of left rect
\node[DefLab] at ($(LhR)+(-0.10,0)$) {$\mathcal{D}$};  % center of right rect

% ================== ISOMORPHISM SYMBOL ==================
\node[font=\Large] at (7.8,0) {$\cong$};

% ================== RIGHT ==================
\coordinate (CR) at (13.0,0);

% symmetry (dashed), same center, slightly larger
\draw[dashed,line width=1pt] (CR) ellipse [x radius=\bx, y radius=\by];

% endpoints on the dashed ellipse (CR): circles with labels INSIDE
\def\wR{8pt} % circle radius for omega markers

\path (CR) ++(-\bx,0) coordinate (CRwL);
\path (CR) ++( \bx,0) coordinate (CRwR);

% (optional) white backing for legibility on top of dashes
\fill[white] (CRwL) circle[radius=\wR];
\fill[white] (CRwR) circle[radius=\wR];

\draw[line width=0.9pt] (CRwL) circle[radius=\wR];
\draw[line width=0.9pt] (CRwR) circle[radius=\wR];

\node[font=\Large] at (CRwL) {$\overline\omega$};
\node[font=\Large] at (CRwR) {${\omega}$};

% Hamiltonian (solid): all blue
\draw[line width=1pt,sptzero] (CR) ellipse [x radius=\ax, y radius=\ay];

% labels inside inner ellipse
\node[Tlab,sptzero] at ($(CR)+(0,\TopY)$) {$\mathcal{T}_0$};
\node[Tlab,sptzero] at ($(CR)+(0,\BotY)$) {$\mathcal{T}_0$};

% symmetry operator label in annulus
\node[SOp] at ($(CR)+(0,\Uy)$) {$g$};

% endpoint blocks on Hamiltonian ONLY (both labeled D')
\path (CR) ++(180:\ax) coordinate (RhL);
\path (CR) ++(  0:\ax) coordinate (RhR);

\draw[fill=gray!10,rounded corners=2pt]
  ($(RhL)+(-0.24,-0.44)$) rectangle ($(RhL)+(0.44,0.44)$);
\draw[fill=gray!10,rounded corners=2pt]
  ($(RhR)+(-0.44,-0.44)$) rectangle ($(RhR)+(0.24,0.44)$);

% labels INSIDE the left/right rectangles (right panel)
\node[DefLab] at ($(RhL)+(0.10,0)$) {$\mathcal{D}'$};
\node[DefLab] at ($(RhR)+(-0.10,0)$) {$\mathcal{D}'$};
\end{tikzpicture}
};

\node at (0,-3){
% bottom row: LEFT has "two spirals = ω(g,h) × (single spiral + D')" ;
% RIGHT is the omega-on-dashed + D' panel, centered to align with CR=13.0
\begin{tikzpicture}[line cap=round,line join=round,scale=0.45,transform shape]

% ---- colors ----
\colorlet{symc}{black!85!black}
\colorlet{defc}{purple!70!black} % <<< same purple
\colorlet{sptzero}{blue!80}

% ---- right-panel geometry copied from main figure ----
\def\ax{3.2} \def\ay{1.9} \def\scl{1.28}
\pgfmathsetmacro\bx{\scl*\ax}
\pgfmathsetmacro\by{\scl*\ay}

% ---- centers to align with main figure ----
\coordinate (CL) at (3.0,0);   % left-center matches main figure's CL
\coordinate (CR) at (13.0,0);  % right-center matches main figure's CR
\def\Yoff{-5.4}                % vertical offset for this row
\def\LeftShift{-2.4}  % <— move left panel this many units left; tweak to taste

% ================== LEFT SMALL PANEL ==================
\begin{scope}[shift={(0,\Yoff)}]
  % spiral params
  \def\a{0.02}
  \def\gap{0}
  \def\Rout{1.5}
  \def\Rin{0.75}
  \def\Rr{0.85}      % RHS spiral scale
  \def\RayL{2.20}    % left defect-ray length
  \def\RayR{1.20}    % RHS defect-ray length
  \def\EqPad{0.35}   % spacing before "= ω(g,h)"
  \def\RHSshift{3.60}% shift from L to start RHS picture

  \coordinate (L) at ($(CL)+(\LeftShift,0)$);

  % two spirals (g,h) centered at endpoint
  \draw[line width=1pt, symc, shift={(L)}]
    plot[samples=160, smooth, domain=\gap:360-\gap, variable=\t]
      ({\Rout*exp(\a*rad(\t))*cos(\t)}, {\Rout*exp(\a*rad(\t))*sin(\t)});
  \draw[line width=1pt, symc, shift={(L)}]
    plot[samples=160, smooth, domain=\gap:360-\gap, variable=\t]
      ({\Rin*exp(\a*rad(\t))*cos(\t)}, {\Rin*exp(\a*rad(\t))*sin(\t)});
  % tiny orientation hints
  \draw[->,symc,line width=1pt, shift={(L)}]
    plot[samples=2, domain=115:130, variable=\t]
      ({\Rout*exp(\a*rad(\t))*cos(\t)}, {\Rout*exp(\a*rad(\t))*sin(\t)});
  \draw[->,symc,line width=1pt, shift={(L)}]
    plot[samples=2, domain=115:130, variable=\t]
      ({\Rin*exp(\a*rad(\t))*cos(\t)}, {\Rin*exp(\a*rad(\t))*sin(\t)});
  % labels inside spirals
  \node[font=\large, symc] at ($(L)+(-\Rout+0.15,0)$) {$g$};
  \node[font=\large, symc] at ($(L)+(-\Rin+0.15,0)$)  {$h$};

  % left defect ray + label
  \draw[defc,line width=1.2pt] (L) -- ($(L)+(\RayL,0)$);
  \node[font=\large, defc] at ($(L)+(\RayL,0.25)$) {$\mathcal{D}'$};

  % equals and phase
  \node[font=\Large, anchor=west] at ($(L)+(\RayL+\EqPad,0)$) {$=\,\omega(g,h)$};

  % ----- RHS-of-left: single spiral gh + short D' ray -----
  \coordinate (LR) at ($(L)+(\RayL+\RHSshift,0)$);
  \draw[line width=1pt, symc, shift={(LR)}]
    plot[samples=160, smooth, domain=\gap:360-\gap, variable=\t]
      ({\Rr*exp(\a*rad(\t))*cos(\t)}, {\Rr*exp(\a*rad(\t))*sin(\t)});
  \draw[->,symc,line width=1pt, shift={(LR)}]
    plot[samples=2, domain=115:130, variable=\t]
      ({\Rr*exp(\a*rad(\t))*cos(\t)}, {\Rr*exp(\a*rad(\t))*sin(\t)});
  \node[font=\large, symc] at ($(LR)+(-\Rr+0.30,0.00)$) {$gh$};

  \draw[defc,line width=1.2pt] (LR) -- ($(LR)+(\RayR,0)$);
  \node[font=\large, defc] at ($(LR)+(\RayR,0.25)$) {$\mathcal{D}'$};
\end{scope}

% ================== RIGHT SMALL PANEL (aligned with CR) ==================
\begin{scope}[shift={(0,\Yoff)}]
  \coordinate (R) at (CR);

  % dashed symmetry ellipse + g
  \draw[dashed,line width=1pt] (R) ellipse [x radius=\bx, y radius=\by];

  % right omega circle (on dashed)
  \def\wR{8pt}
  \path (R) ++(\bx,0) coordinate (RwR);
  \fill[white] (RwR) circle[radius=\wR];
  \draw[line width=0.9pt] (RwR) circle[radius=\wR];
  \node[font=\Large] at (RwR) {$\omega$};

  % inner solid Hamiltonian ellipse (all blue)
  \draw[line width=1pt, sptzero] (R) ellipse [x radius=\ax, y radius=\ay];

  % labels inside
  \node[font=\Large, sptzero] at ($(R)+(0,1.06)$) {$\mathcal{T}_0$};
  \node[font=\Large, sptzero] at ($(R)+(0,-1.06)$) {$\mathcal{T}_0$};

  % symmetry label in annulus
  \node[font=\Large] at ($(R)+(0,{-\ay - 0.30})$) {$g$};

  % right D' block on Hamiltonian
  \path (R) ++(0:\ax) coordinate (RhR);
  \draw[fill=gray!10,rounded corners=2pt]
    ($(RhR)+(-0.44,-0.44)$) rectangle ($(RhR)+(0.24,0.44)$);
  \node[font=\Large, text=defc] at ($(RhR)+(-0.10,0)$) {$\mathcal{D}'$}; % <<< purple
\end{scope}

\end{tikzpicture}
};

% ================== THIRD ROW: (c) ==================
\node at (0,-6.0){
\begin{tikzpicture}[x=1cm,y=1cm,line cap=round,line join=round,>=stealth]
  % colors
  \colorlet{defc}{violet!90!black}   % defect line (unchanged)
  \colorlet{gc}{blue!75!black}       % blue
  \colorlet{Ac}{blue!75!black}       % all lines blue

  % styles (PRL column)
  \tikzset{
    defline/.style={defc, line width=1.2pt},
    gline/.style={gc,   line width=1.2pt},
    Aline/.style={Ac,   line width=1.2pt},
    blab/.style={text=Ac, font=\footnotesize}, % labels in blue
    glab/.style={text=gc, font=\footnotesize},
    dlab/.style={text=defc, font=\footnotesize}
  }

  % compact geometry
  \def\sep{4.6}
  \def\H{0.95}
  \def\R{1.55}
  \def\yoff{0.42}
  \def\xsplit{\R-0.75}

  % ---------------- NEW LEFT PANEL ----------------
  \begin{scope}[shift={(-\sep/2,0)}]
    % background tints (both blue)
    \fill[blue!6] (-\R,-\H) rectangle (0,\H);
    \fill[blue!6]  (0,-\H)  rectangle (\R,\H);

    % vertical defect
    \draw[defline] (0,-\H) -- (0,\H);
    \node[dlab, anchor=north west] at (0.05,-\H) {$\mathcal{D}'$};

    % straight horizontals (rho_0)
    \draw[Aline] (0,\yoff) -- node[blab, pos=0.58, above] {$\rho_0(g)$} (\R,\yoff);
    \draw[Aline] (0,-\yoff) -- node[blab, pos=0.58, below] {$\rho_0(h)$} (\R,-\yoff);

    % blue curves fusing to rho_0(gh); labels ON the curves
    \coordinate (M) at (-0.75,0);
    \draw[gline] (0,\yoff)
      .. controls (-0.35,\yoff) and (-0.55,0.30) ..
      node[glab, pos=0.45, above] {$\rho_0(g)$}
      (M);
    \draw[gline] (0,-\yoff)
      .. controls (-0.35,-\yoff) and (-0.55,-0.30) ..
      node[glab, pos=0.45, below] {$\rho_0(h)$}
      (M);
    \draw[gline] (-\R,0) -- node[glab, pos=0.55, above] {$\rho_0(gh)$} (M);
  \end{scope}

  % equals sign with omega(g,h)
  \node[font=\small] at (0,0) {$=\,\omega(g,h)$};

  % ---------------- NEW RIGHT PANEL  ----------------
  \begin{scope}[shift={( \sep/2,0)}]
    % background tints (both blue)
    \fill[blue!6] (-\R,-\H) rectangle (0,\H);
    \fill[blue!6] (0,-\H)  rectangle (\R,\H);

    % vertical defect
    \draw[defline] (0,-\H) -- (0,\H);
    \node[dlab, anchor=north west] at (0.05,-\H) {$\mathcal{D}'$};

    % blue rho_0(gh) on the line itself
    \draw[gline] (-\R,0) -- node[glab, pos=0.35, above] {$\rho_0(gh)$} (0,0);

    % trunk (rho_0)
    \draw[Aline] (0,0) -- node[blab, pos=0.6, above] {$\rho_0(gh)$} (\xsplit,0);

    % upper leg (rho_0(g))
    \draw[Aline] (\xsplit,0)
      .. controls (\xsplit+0.35, 0.28) and (\R-0.35, \yoff) ..
      node[blab, pos=0.62, above] {$\rho_0(g)$}
      (\R,\yoff);

    % lower leg (rho_0(h))
    \draw[Aline] (\xsplit,0)
      .. controls (\xsplit+0.35,-0.28) and (\R-0.35,-\yoff) ..
      node[blab, pos=0.62, below] {$\rho_0(h)$}
      (\R,-\yoff);
  \end{scope}
\end{tikzpicture}
};

% row labels
\node at (-4,1){\textbf{(a)}};
\node at (-4,-1.8){\textbf{(b)}};
\node at (-4,-4.9){\textbf{(c)}};

\end{tikzpicture}

\caption{\textbf{Symmetric interfaces and defect anomalies.}
We consider the case where theories $\mathcal{T}_0$ and $\mathcal{T}_1$ are related by an SPT entangler, which holds for unitary internal symmetries where local charge assignments agree.
\textbf{(a)} Acting with a half-chain SPT entangler turns a $\mathcal{T}_0$–$\mathcal{T}_1$ ring with interfaces  $\mathcal{D}$ into a uniform $\mathcal{T}_0$ ring with two localized defects $\mathcal{D}'$ and a \emph{modified} symmetry $U'_g$ (dashed line) that carries local projective factors $\omega\in H^2(G,U(1))$ at the defect sites.
\textbf{(b)} A single $\mathcal{D}'$ must be $G$-symmetric with respect to that projective action, enforcing a defect anomaly constraint. The defect anomaly constraint on $\mathcal{D}'$ is a necessary and sufficient condition for any admissible interface $\mathcal{D}$.
\textbf{(c)} In the IR, the entangler action shifts the phases of $G$-network tri-junctions by $\omega$, rendering the phase assignments on the right to match those on the left so that it is appropriate to describe $\mathcal{D}'$  as a \textit{defect} within a CFT. 
The projective phase in Fig.~\ref{fig:G-symmetric-defect}(b) signals the defect anomaly of $\mathcal{D}'$.} 
\label{fig:entangler-defects}
\end{figure}
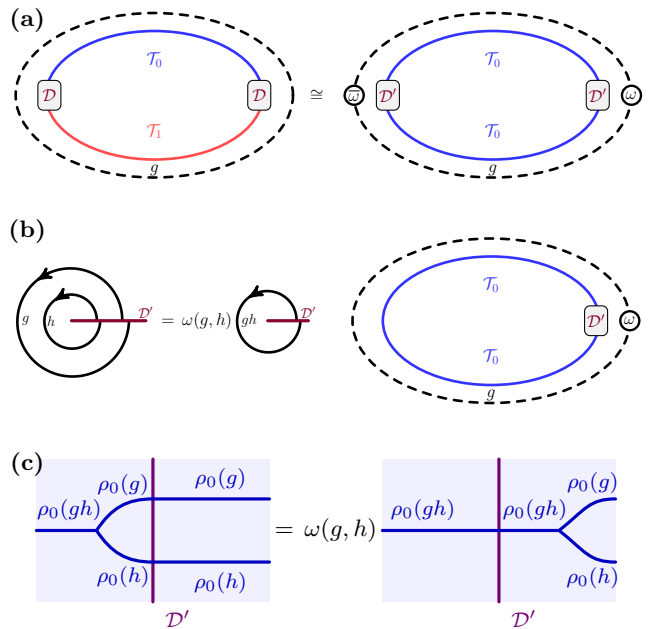

We now focus on pairs of critical theories related by stacking a conventional gapped SPT phase or, equivalently, by conjugation with a local unitary  “SPT entangler” $U_{\rm ent}$~\cite{Chen_2013,Chen_2014,Else_2014,Bultinck:2019zzo,pivot,zhang2022topologicalinvariantssptentanglers, prembabu2025multicriticalitypurelygaplessspt}:
\[
H_1 \;=\; U_{\rm ent}\, H_0\, U_{\rm ent}^\dagger .
\]

For unitary, non-anomalous $G$ which acts faithfully on the low-energy theory, we expect two distinct SECs with the same local operator charges to differ by such a gapped SPT classified by $H^2(G,U(1))$. 
This is because such SECs are in one-to-one correspondence with different ways of gauging (or orbifolding) $G$, which are known to be classified by  discrete torsion classes given by $H^2(G,U(1))$~\cite{Vafa_1986}.
As is known from the study of gapped SPTs, the entangler $U_{\rm ent}$ can be chosen to globally commute with the group $G$, but it cannot be built out of $G$-symmetric gates \cite{Else_2014}. This means that it can act projectively on nonlocal operators, i.e., it can change the symmetry action on string operators.

Here we demonstrate that the \emph{conformal interface} inherits this projective data in a precise way, leading to a gapless generalization of the conventional notion of SPT edge anomalies. In particular we show:

\medskip

\noindent
\fbox{
\parbox{\dimexpr\linewidth-2\fboxsep-2\fboxrule}{
\textbf{Claim.} The allowed $G$-symmetric conformal interfaces between $\mathcal{T}_0$ and $\mathcal{T}_1 \equiv \mathcal{T}_0\times \textrm{$G$-SPT}$ are precisely those conformal defects whose endpoint Hilbert space realizes the projective representation of the $G$-SPT.
}
}

\medskip

The argument is summarized in Fig.~\ref{fig:entangler-defects}.
On a periodic chain, we can set up $\mathcal{T}_0$ on one half and $\mathcal{T}_1$ on the other, and near each of the two interfaces allow arbitrary local $G$-symmetric couplings which we call $\mathcal{D}$, whose IR fate we seek to understand. We can then act with the \textit{truncated} SPT entangler only on the $\mathcal{T}_1$ half. This maps the system to a uniform $\mathcal{T}_0$ chain with two localized defects $\mathcal{D}'$ at the former interfaces, while the symmetry becomes
\[
U'_g \;\equiv\; U_{\rm ent}^{\rm half}\, U_g\, U_{\rm ent}^{\rm half\dagger}.
\]
Crucially, $U'_g$ agrees with $U_g$ away from the defects but acquires local decorations near the defect sites that realize the associated \textit{projective} symmetry at each of the defects.  A localized defect $\mathcal{D}'$ of $\mathcal{T}_0$ is therefore admissible exactly when it is symmetric with respect to this projective action.

To concretely relate this symmetry condition to the endpoint Hilbert space, we may consider a periodic $\mathcal{T}_0$ chain with a \textit{single} insertion of $\mathcal{D}'$ (Fig.~\ref{fig:entangler-defects}(b)). In this setup, the global symmetry action itself is projective; hence the model's low energy spectrum, which is given by the endpoint Hilbert space of $\mathcal{D}'$, realizes the projective representation. 
Thus the admissible defect classification reduces to intrinsic CFT data: $\mathcal{D}'$ is admissible if and only if its endpoint Hilbert space carries the required projective representation. Conversely, any such $\mathcal{D}'$ pulls back to a $G$-symmetric interface between $\mathcal{T}_0$ and $\mathcal{T}_1$ by applying $(U_{\rm ent}^{\rm half})^\dagger$.

This endpoint property is the “defect  't Hooft anomaly” discussed in recent literature~\cite{Antinucci:2024izg,Komargodski:2025jbu,Shao:2025qvf} (see also \cite{Delmastro:2022pfo,Brennan:2022tyl,Brennan:2025acl}). In gapped settings (where the same argument applies), the anomaly is saturated by the familiar degenerate SPT edge modes living in a decoupled Hilbert space.
Although the same can occur in gapless settings (the ``screened DQFT'' of Ref.~\onlinecite{Antinucci:2024izg}), there are also richer possibilities in which genuinely gapless defect degrees of freedom themselves carry the anomaly, encoding an intrinsic structure of the conformal defect.
Stable examples often realize the anomaly through degenerate, SSB-like sums of conformal defects, though this need not always be the case.

We stress that although a ring with the  insertion of a single defect with nontrivial defect anomaly is necessarily degenerate, a ring with \emph{two} defect insertions may be nondegenerate. 
An example is the Kramers-Wannier defect in the Ising CFT, which has a $\mathbb{Z}_2\times \mathbb{Z}_2^T$ defect  anomaly \cite{Seiberg:2024gek}. The spectrum on a ring with a single insertion of this defect is two-fold degenerate because of the projective representation of $\mathbb{Z}_2\times \mathbb{Z}_2^T$. 
However, on a ring with two such defects, the anomalies cancel, and  they can be fused to the sum of the identity defect and an invertible $\mathbb{Z}_2$ defect, giving rise to a non-degenerate spectrum. 
This is consistent with the examples in Section~\ref{sec:ising}.

\subsection{CFTs which  absorb SPTs}

If the theory $\mathcal{T}_0$ is invariant under the SPT entangler (so that $\mathcal{T}_0=\mathcal{T}_1$), our setup does not produce an interface between \emph{distinct} gapless SPTs. 
Nevertheless, the same reasoning about defect anomaly applies. 
In this case one may take $\mathcal{D}'$ to be the \textit{invertible} entangler defect, which is anomalous.
(The corresponding interface $\mathcal{D}$ is therefore the identity interface.) 
Moreover, such an invertible anomalous defect exists if and only if the entangler is a symmetry  with a certain type of 't Hooft anomaly of the low-energy CFT. 
The \emph{presence} or \emph{absence} of an invertible, anomalous defect serves as a diagnostic for whether two sides are distinguished as gapless SPTs.

As a concrete example, consider the spin chain with Hamiltonian \cite{LevinGu}
\begin{equation}
H= -\sum_j\!\big(X_j+Z_{j-1}X_jZ_{j+1}\big). 
\end{equation}
It is invariant under \(G=\mathbb{Z}_2\!\times\!\mathbb{Z}_2\) generated by \(\prod_{j\,\text{even}}X_j\) and \(\prod_{j\,\text{odd}}X_j\). 
The $G$-entangler is \(U_{\rm ent}=\prod_j \mathrm{CZ}_{j,j+1}\), which leaves this Hamiltonian invariant and generates another $\mathbb{Z}_2$ symmetry. Acting with \(U_{\rm ent}\) on a half chain (sites \(j\ge0\)) leaves the bulk away from the cut unchanged but modifies the two junctions. 
The system with a single such defect is \cite{Seiberg:2025bqy,Cheng:2022sgb}
\begin{equation}
-\,Z_{-1}X_0 \;-\; X_0Z_1 \;-\! \sum_{j\neq 0}\!\big(X_j+Z_{j-1}X_jZ_{j+1}\big).
\end{equation}
This is the invertible defect for the entangler $\mathbb{Z}_2$ symmetry.
The $G=\mathbb{Z}_2\times \mathbb{Z}_2$ symmetry is realized projectively in the presence of this invertible defect, signaling a defect anomaly. 
More specifically, in the presence of this defect, 
the symmetry $G$ is now generated by the following modified operators
$\prod_j X_{2j}$ and $Z_0 \prod_j X_{2j-1}$, which anticommute with each other. 
 This in turn is related to the type-III 't Hooft anomaly \cite{Seifnashri:2024dsd,Pace:2024oys,Seiberg:2025bqy} of the full $\mathbb{Z}_2\times \mathbb{Z}_2\times \mathbb{Z}_2$ symmetry including the $\prod_j \mathrm{CZ}_{j,j+1}$ generator. 
 (See Refs.~\onlinecite{deWildPropitius:1995cf,Wang:2014tia,Yoshida16} for the 2+1d SPT phase for this anomaly.)

\subsection{Classification for Ising case} \label{subsec:projective_ising}

As an example, we illustrate how the above general result can be used to classify the possible interfaces between critical Ising chains. A concrete instance of the defect anomaly arises in the preceding example discussed in Section~\ref{sec:ising}: \(H_{\rm Ising}\) and \(H_{\rm Ising}^\mu\) differ in twisted-sector charges and are related by a \(\mathbb{Z}_2 \times \mathbb{Z}_2^T\) SPT entangler $\prod_n {\rm CZ}_{n,n+1}$. 
Acting this entangler on the half chain \([j_L,j_R]\) with the \(H_{\rm Ising}^\mu\) region turns the Hamiltonian into the ordinary transverse-field Ising Hamiltonian except around the two sites $j_L$ and $j_R$. 
The  symmetry generators become \(Z_{j_L} Z_{j_R} \prod_j X_j\) and complex conjugation \(\mathcal{K}\),
 which act projectively at the two defect sites \(j_L\) and \(j_R\).   
Hence the classification reduces to finding defects with a defect anomaly under $\mathbb{Z}_2\times \mathbb{Z}_2^T$. Among nondegenerate Ising defects, the only ones with a degenerate endpoint Hilbert space are the Neumann defects with \(g=\sqrt{2}\); the other degenerate examples are the \(g=1\) spontaneously aligned/anti-aligned defects and unstable higher-\(g\) direct-sum realizations. 
This shows that our interfaces in Fig.~\ref{fig:Ising_interface}(b) captures all options of codimension two or less in the phase diagram.

The Kramers–Wannier transformation maps \(H_{\rm Ising}^\mu \to H_{\rm Ising}^\sigma\) while leaving \(H_{\rm Ising}\) invariant, so the same criterion classifies the interfaces between \(H_{\rm Ising}^\sigma\) and \(H_{\rm Ising}\), thereby capturing Fig.~\ref{fig:Ising_interface}(a). In Appendix~\ref{app:Hsimga_Hmu_defects_proof} we further show how defect anomalies constrain the interface between \(H_{\rm Ising}^\sigma\) and \(H_{\rm Ising}^\mu\).

\section{Higher dimensions \label{sec:higher}}

Thus far, we have focused on 1+1d critical chains since that is the setting where CFTs are best controlled. However, we expect the key insights to generalize to higher dimensions. The concept of `symmetry-enriched criticality' is known to generalize beyond 1+1d \cite{Scaffidi_2017}, including the case where the symmetry acts faithfully in the low-energy theory \cite{Verresen_2021}. In many instances, these CFTs can again by distinguished by the charges of operators, or by twisted partition functions. The intuition we have built up in this work thus far, suggests that the spatial interface between such distinct symmetry-enriched theories cannot be a topological invertible defect, i.e., cannot be transparent. For instance, in the special case of CFTs related by SPT-entanglers, Section~\ref{sec:projective} tells us that the allowed interfaces between 2+1d CFTs are classified by studying anomalous 1+1d defects. While this excludes a trivial `transparent' defect, it would be interesting for future work to investigate the resulting interface phase diagrams, similar to our examples in Section~\ref{sec:ising}.

One of the simplest examples is given by the 2+1d Ising CFT. If we fix our $\mathbb Z_2$ symmetry at the UV to be $\prod_n X_n$, there are, in fact, \emph{two} distinct symmetry-enriched versions of the Ising CFT. The first is the phase transition between the trivial paramagnet and the ferromagnet, whereas the second is a transition between the same ferromagnet and a non-trivial gapped SPT phase protected by $\mathbb Z_2$ symmetry \cite{LevinGu}. Note that according to the path-based definition, these must be in distinct phases, since at each point along such a path in parameter space there should be a unique answer to the question ``\emph{in what SPT phase is the nearby symmetry-preserving gapped phase?}''. Such distinct symmetry-enriched CFTs can be probed with twisted partition functions: in Ref.~\onlinecite{Verresen_2021} this was numerically demonstrated for the torus partition function of two distinct 2+1d Ising$^{3}$ CFTs with $\mathbb Z_2^3$ symmetry, and recently in Ref.~\onlinecite{Vibhu25} for Klein bottle partition function of the two 2+1d Ising CFTs with $\mathbb Z_2$ symmetry.

As concrete lattice realizations of these two Ising CFTs, first consider the usual transverse-field Ising model on the triangular lattice:
\begin{equation}
H_0 \;=\; -J \sum_{\langle j k\rangle} Z_j Z_k \;- \sum_{j} X_j.
\label{eq:H0_2d}
\end{equation}
This flows to the 2+1d Ising CFT for $J \approx 0.20973$ \cite{Bloete02}. To obtain the phase transition between a ferromagnet and a non-trivial $\mathbb Z_2$ SPT \cite{LevinGu,Yoshida16}, we can consider
\begin{equation}
H_1 = U_{\rm CCZ} H_0 U_{\rm CCZ}^\dagger \quad \textrm{with } U_{\rm CCZ} = \prod_{\langle i jk \rangle } {\rm CCZ}_{ijk},
\end{equation}
where the product over $\langle ijk\rangle$ ranges over all 
triangular plaquettes of the triangular lattice, and ${\rm CCZ}_{ijk}$ assigns a phase factor of $-1$ if and only if all spins on a given triangle point down.

A spatial interface between $H_0$ and $H_1$ will be non-trivial, even if both are at their Ising critical point $J \approx 0.20973$. For instance, applying the technique of Section~\ref{sec:projective}, we can apply the SPT (dis)entangler on half of space, such that we obtain a defect line of the usual transverse-field Ising model $H_0$. In particular, the $\mathbb Z_2$ symmetry is now 
\begin{equation}\label{CZX}
\prod_n X_n \times \prod_{ \langle m,n \rangle \in {\rm line}} \mathrm{CZ}_{m,n}.
\end{equation} 
Along the defect line, the $\mathbb{Z}_2$ symmetry is generated by $\prod_j X_j \prod_j \mathrm{CZ}_{j,j+1}$  \cite{Chen:2011bcp,LevinGu}, which is known to have an 't Hooft anomaly in 1+1d classified by $H^3( \mathbb{Z}_2,U(1))=\mathbb{Z}_2$. 

 One 2+1d Hamiltonian invariant under \eqref{CZX} is
\begin{equation}
H' = -J \sum_{\langle jk\rangle} Z_j Z_k - \sum_{j \notin {\rm line}} X_j.
\end{equation}
This defect line clearly spontaneously breaks the $\mathbb Z_2$ symmetry, since $Z_j$ commutes with $H'$ for $j$ along the defect line. 
We expect that this is the generic defect line for this symmetry. By tuning a parameter, one could conceivably access new types of critical defect lines. One way of setting this up is by coupling a 1+1d compact boson CFT with anomalous $\mathbb Z_2$ symmetry $\phi \to \phi+\pi$ and $\theta \to \theta + \pi$ to the 2+1d Ising CFT in a $\mathbb Z_2$-symmetric fashion, which allows for couplings of the form $\sigma  \cos \phi$, where $\sigma$ is the spin operator of the 2+1d Ising CFT and $\cos \phi$ is a vertex operator of the 1+1d compact boson CFT. We leave a study of the resulting defect phase diagram to future work. 
Defects of 2+1d Ising CFTs have been studied in Refs.~\onlinecite{Diehl_1997,Liendo:2012hy,Billo:2013jda,Gaiotto:2013nva,Gliozzi:2015qsa,Yamaguchi:2016pbj,Liendo:2019jpu,Zhou:2024dbt,Cuomo:2024psk,Lanzetta:2025xfw}.

\section{Outlook}

In this work we explore the fundamental question of what interfaces may exist between two symmetry-enriched critical theories, and conversely what the admissible interfaces tell us about signatures of the symmetry-enriched bulk. We find various constraints, including a general rule that any two symmetry-enriched criticalities with distinct charge assignments, either local or nonlocal, can be spatially separated only by a non-invertible defect. In simple cases of the 1+1d Ising CFT enriched by $\mathbb{Z}_2\times \mathbb{Z}_2^T$ symmetry, we find that stable interfaces are factorizing and sometimes display symmetry-protected degenerate modes. Finally, in analogy to conventional SPTs, we show that interfaces between critical theories which differ by gapped SPT entanglers can be characterized by defect anomalies. 
This approach also applies to higher-dimensional examples.

This work promotes an interface-based view of symmetry-enriched criticality, characterizing gapless phases by the interfaces they admit. Many questions remain for more general conformal field theories. In the 1+1d Ising CFT examples, we found that the most stable interfaces are factorizing. It would be interesting to investigate whether this holds more generally; this would resemble recent results on pinning-field defects \cite{Popov:2025cha}. 
Moreover, in the Ising examples, the fate of the interface seemed tied to the nearby phase diagram. For example, $H_{\rm Ising}$ and $H_{\rm Ising}^\sigma$ share an adjacent trivial paramagnetic phase and a stable nondegenerate $\ket{f}\bra{f}$ interface, while $H_{\rm Ising}$ and $H_{\rm Ising}^\mu$ share an adjacent ferromagnetic phase with $\mathbb{Z}_2^T$-even order parameter and a $\mathbb{Z}_2$ spontaneous-symmetry-breaking stable interface. Intuitively it makes sense that one can smoothly interpolate a relevant perturbation into a gapped phase between the critical theories and contract the gapped region. Understanding whether this relation holds more generally is a promising line of work.\footnote{We thank Ryan Thorngren for stimulating discussions on this point.}

It would be interesting to further map out the space of $G$-symmetric interfaces between two symmetry-enriched criticalities, beyond the examples discussed in this work. A key problem is to understand defects with defect anomalies; we have shown this is equivalent to classifying the interfaces between gapless SPTs distinguished by entanglers. For general CFTs, bootstrap methods may constrain this family.  
An exciting case study would be given by the 2+1d example we proposed in Section~\ref{sec:higher}. More generally, our work opens up the possibility that for gapless phases, the bulk–boundary correspondence might have to be replaced by a `bulk-defect' or `bulk-interface' correspondence. 

\begin{acknowledgements}
The authors thank Ehud Altman, Takamasa Ando, Andrea Antinucci, Yichul Choi, Maine Christos, Dominic Else, Wenjie Ji, Nick Jones, Zohar Komargodski, Ryan Lanzetta, Maxim Metlitski, Masaki Oshikawa, Ingo Runkel, Sakura Schäfer-Nameki, Nathan Seiberg, Sahand Seifnashri, Madhav Sinha, Ryan Thorngren, Frank Verstraete, Chong Wang, Zack Weinstein, and Yunqin Zheng for insightful discussions, with special thanks to Yifan Wang for an inspiring discussion and encouragement at early stages of this project. S.P. thanks P. Chandrasekar for support. 
S.H.S. is supported by the Simons Collaboration on Ultra-Quantum Matter, which is a grant from the Simons Foundation (651444, SHS). S.H.S. is also supported in part by
NSF grant PHY-2449936. 
S.P. was supported by the National Science Foundation grant NSF DMR-2220703.  This research was supported in part by grant NSF PHY-2309135 to the Kavli Institute for Theoretical Physics (KITP) while attending the `Generalized Symmetries' program.
\end{acknowledgements}

 \bibliographystyle{ytphys}

 \bibliography{references}  

\appendix

\section{Interfaces between gapped SPT phases}\label{app:gappedSPT}

Although the focus in the main text is on \emph{gapless} theories, our results also apply to  \emph{gapped} theories as special cases. 
This corresponds to  choosing the bulk to be a trivial CFT whose only local operator is the identity. 
Even though that case is already well-understood, it can be useful to briefly revisit some of our key concepts and results in that case, as a conceptual anchor point.

As a simple instance, we will consider gapped SPT phases protected by a \(G=\mathbb{Z}_2\times\mathbb{Z}_2\) symmetry generated by
\begin{equation}\label{Z2Z2}
\prod_j X_{2j}~~\text{and}~~\prod_j X_{2j+1}.
\end{equation}
The trivial phase is given by $H_0 = - \sum_n X_n$, and the non-trivial SPT phase is given by the cluster model \cite{Briegel_2001,Son_2011}, 
\begin{equation}
H_1 = -\sum_n Z_{n-1} X_n Z_{n+1}.
\end{equation}
The two Hamiltonians are related by the entangler $U = \prod_n \mathrm{CZ}_{n,n+1}$ as  $H_1 = U H_0 U^\dagger$.

\subsection{Non-invertible symmetric interface}

Consider an interface between $H_0$ and $H_1$ on a ring of  $N$ sites (with $N=0$ mod 4):
\begin{equation}
H = - \sum_{j=1}^{\frac N2-1} X_j - \sum_{j=\frac N2+1}^{N-1} Z_{j-1} X_j Z_{j+1}. \label{eq:gapped}
\end{equation}
which preserves the $\mathbb{Z}_2\times\mathbb{Z}_2$ symmetry. 
Our general result in Section~\ref{sec:noninvertible} then implies the two interfaces have to be non-invertible. 
When the bulk is trivially gapped, a non-invertible interface simply means one with degenerate modes. 
To see the degeneracy, we note that the Hamiltonian is a stabilizer where every term commutes with each other. 
Furthermore, it commutes with  $Z_{\frac N2}$ and $X_{\frac N2} Z_{\frac N2+1}$ at one interface, as well as  $Z_N$ and $Z_{N-1} X_N$ at the other interface. 
These two pairs of  anticommuting operators imply a four-fold degeneracy, confirming our  claim above.

We can equivalently view this interface system as the cluster Hamiltonian $H_1$ on an open chain. 
Then the four-fold degeneracy corresponds to the familiar symmetry-protected edge modes in the discussion of gapped SPT phases. 
While the reformulation of these edge modes in terms of non-invertible interfaces may unnecessarily fancy, it proves to be the correct generalization when extended to the gapless setting as we argued in Section~\ref{sec:noninvertible}. 

\subsection{Invertible non-symmetric interface}

Next, consider a different interface:
\begin{equation}
H = - \sum_{j=1}^{\frac N2-1} X_j - \sum_{j=\frac N2}^{N-1} Z_{j-1} X_j Z_{j+1}
- X_{\frac N 2-1}Z_{\frac N2} - Z_{N-1}X_N.
\end{equation}
This is a stabilizer Hamiltonian with $N$ terms. The ground state is therefore unique on a ring, implying that the two interfaces must be invertible. 
However, this interface Hamiltonian does not commute with the $\mathbb{Z}_2\times\mathbb{Z}_2$ symmetry generated by the operators in \eqref{Z2Z2}.\footnote{Having said that, this interface Hamiltonian does preserves a different $\mathbb{Z}_2\times \mathbb{Z}_2$ symmetry generated by $Z_{\frac N2-1}\prod_j X_{2j}$ and $Z_N\prod_j X_{2j+1}$. This is not `on-site', in the sense that it is not a tensor product representation of local linear $\mathbb Z_2 \times \mathbb Z_2$ representations.} 
Hence, this is an invertible, but not symmetric interface, again consistent with the result in Section~\ref{sec:noninvertible}.

See, for example, the appendices of Ref.~\onlinecite{Seifnashri:2024dsd} for more discussions of these two interface Hamiltonians above.

\subsection{Defect  't Hooft anomaly}

Starting with the symmetric, non-invertible interfaces in Eq.~\eqref{eq:gapped}, we can apply the SPT entangler to half the system corresponding to the cluster Hamiltonian as in  Fig~\ref{fig:entangler-defects} in Section~\ref{sec:projective}. 
Then the Hamiltonian simply becomes
\begin{equation}
H' = - \sum_{j \notin \{\frac N2,N\} } X_j.
\end{equation}
This is simply the trivial Hamiltonian with two sites $\frac N2$ and $N$ missing, giving rise to the four-fold degeneracy mentioned above. 
The two symmetry operators in \eqref{Z2Z2} become
\begin{equation}
\prod_{j} X_{2j} \quad \textrm{and} \quad Z_{\frac N2} Z_N \prod_j X_{2j+1}. 
\end{equation}

Now that we have turned an interface between two gapped phases into a defect within the same trivially gapped phase, we can  consider the configuration with a single defect on a ring
\begin{equation}
H' = - \sum_{j \neq {\frac N2}} X_j,
\end{equation}
with symmetry generators 
\begin{equation}
\prod_j X_{2j}~~~\text{and}~~~Z_{\frac N2} \prod_j X_{2j+1}.
\end{equation}
It is now clear that they form a projective representation of the $\mathbb Z_2 \times \mathbb Z_2$ symmetry, in agreement with our general claim that a single defect realizes an anomalous symmetry action.

\section{Ising defects, defect spectra, and RG flows}
\label{app:Affleck_Oshikawa_defect_classification}
\begin{figure}
    \centering
\begin{tikzpicture}[
  x=3cm, y=3cm,
  baseline=(current bounding box.south),
  seg/.style={very thick},
  lab/.style={font=\small, inner sep=1pt},
  dot/.style={circle, fill=black, inner sep=0pt, minimum size=2pt},
  arr/.style={->, thin}
]
\def\nheight{0.5}
\def\tick{0.06} % half tick length (in y-units)

% --- segments ---
\draw[seg, red] (0,0) -- (2,0);              % D(phi) line (0 .. π)
\draw[seg, blue] (0,\nheight) -- (1,\nheight); % N(phi) line (0 .. π/2)

% --- y=0 dots + labels ---
\fill[dot] (0,0) circle[radius=1.1pt];
\node[lab, above left=2pt]  at (0,0) {$\uparrow\uparrow + \downarrow\downarrow$};

\fill[dot] (0.5,0) circle[radius=1.1pt];
\node[lab, above=2pt]       at (0.5,0) {identity};

\fill[dot] (1.0,0) circle[radius=1.1pt];
\node[lab, above=2pt]       at (1.0,0) {$ff$};

\fill[dot] (1.5,0) circle[radius=1.1pt];
\node[lab, above=2pt]       at (1.5,0) {$\mathbb{Z}_2$ defect};

\fill[dot] (2,0) circle[radius=1.1pt];
\node[lab, above right=2pt] at (2,0)   {$\uparrow\downarrow + \downarrow\uparrow$};

% --- y=\nheight dots + labels ---
\fill[dot] (0,\nheight) circle[radius=1.1pt];
\node[lab, above left=2pt]  at (0,\nheight) {$f\uparrow + f\downarrow$};

\fill[dot] (0.5,\nheight) circle[radius=1.1pt];
\node[lab, above=2pt]       at (0.5,\nheight) {KW duality};

\fill[dot] (1,\nheight) circle[radius=1.1pt];
\node[lab, above right=2pt] at (1,\nheight) {$\uparrow f + \downarrow f$};

\node[lab, right=8pt, blue] (ANNN) at (1.6,\nheight) {$N(\phi),\ g{=}\sqrt{2}$};
\draw[arr, blue] (ANNN.west) to[out=180,in=-25] (1.1,\nheight-0.01);

\node[lab, right=8pt, red] (ANND) at (1.6,\nheight*0.6) {$D(\phi),\ g{=}1$};
\draw[arr, red] (ANND.west) to[out=180,in=45] (1.25,0.15);

% --- angle ticks: TOP line 0 .. π/2 (labels below to avoid KW) ---
\foreach \x/\t in {0/{$0$}, 0.5/{$\pi/4$}, 1/{$\pi/2$}}{
  \draw (\x,\nheight-\tick) -- (\x,\nheight+\tick);
  \node[lab, below=4pt] at (\x,\nheight) {\t};
}

% --- angle ticks: BOTTOM line 0 .. π (alternate label sides to avoid overlaps) ---
% below: 0, π/2, π
\foreach \x/\t in {0/{$0$}, 1/{$\pi/2$}, 2/{$\pi$}}{
  \draw (\x,0-\tick) -- (\x,0+\tick);
  \node[lab, below=4pt] at (\x,0) {\t};
}
% above: π/4, 3π/4
\foreach \x/\t in {0.5/{$\pi/4$}, 1.5/{$3\pi/4$}}{
  \draw (\x,0-\tick) -- (\x,0+\tick);
  \node[lab, below=3pt] at (\x,0) {\t};
}

\end{tikzpicture}

    \caption{A full classification of the nondegenerate elementary defect moduli spaces of the Ising model, adapted from Fig.~4 of Ref.~\onlinecite{Kormos_2009}. The two continuous moduli spaces of nondegenerate defects are the Dirichlet defects $D(0<\phi<\pi)$ and the Neumann defects $N(0<\phi<\pi/2)$. At the endpoints of these spaces are degenerate defects, the $\mathbb{Z}_2$ SSB sums of two decoupled Ising boundaries. The topological and factorizing defects are labeled.}
    \label{fig:Ising-all-defects}
\end{figure}
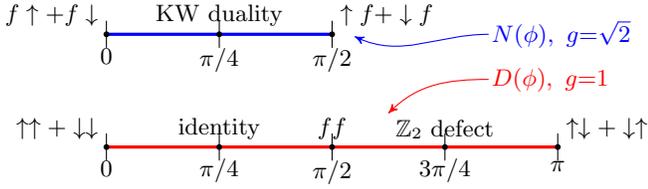

We review the fully classified defects of the  Ising CFT as well as their spectra \cite{Kormos_2009, Oshikawa:1996dj}. The simple defects (those that are not a direct sum of other defects) are shown in Fig. \ref{fig:Ising-all-defects}.

There are two continuous one-parameter families of elementary defects.
Dirichlet defects \(D(\phi)\), with \(0\le \phi \le \pi\), all have Affleck--Ludwig entropy \(g=1\).
This family contains the identity defect \(D(\tfrac{\pi}{4})\) and the invertible topological \(\mathbb{Z}_2\)  defect \(D(\tfrac{3\pi}{4})\).
It also passes through three special, factorizing points corresponding to Ising boundary conditions: the \( \ket{f}\bra{f}\) defect \(D(\tfrac{\pi}{2})\), the spontaneously aligned fixed defect \(D(0)\), and the spontaneously anti-aligned fixed defect \(D(\pi)\).
The last two are 
degenerate; they spontaneously break the \(\mathbb{Z}_2\) symmetry and decompose into degenerate sums of two fixed defects, each with \(g=\tfrac{1}{2}\).
Neumann defects \(N(\phi)\), with \(0\le \phi \le \tfrac{\pi}{2}\), are obtained by fusing the non-invertible Kramers-Wannier duality defect to the left of \(D(\phi)\).
In addition, finite direct sums with integer multiplicities of the above defects exist.

The spectrum of the Ising CFT on a circle of length $L$ with two diametrically opposite Dirichlet defects \(D(\phi_1)\) and \(D(\phi_2)\) is captured by the partition function, which is a function of $q \equiv e^{-2\pi \beta/L}$ with $\beta$ being the inverse temperature. 
We can further restrict the partition function to a fixed $\mathbb{Z}_2$ global symmetry charge sector $(s=0,1)$, under the untwisted $(t=0)$ or the  $\mathbb{Z}_2$-twisted $(t=1)$ boundary condition. Explicitly, these partition functions are 
\begin{equation}
Z^{s,t}_{D(\phi_1),D(\phi_2)}(q)
= \frac{1}{\eta(q)}
  \sum_{n\in\mathbb{Z}}
  q^{2\left(n+\frac{t}{2}+\frac{\phi_1 - (-1)^{s+t} \phi_2}{2\pi}\right)^2},
\end{equation}
where $\eta(q)$ is the Dedekind eta function. 
These formulas also cover the special case a single Dirichlet defect insertion, by setting $\phi_2 = \pi/4$.

The spectrum with a single Neumann defect is independent of \(\phi\), \(s\), or \(t\):
\begin{equation}
Z^{s,t}_{N(\phi)} = \chi_{1/16}(q) \left( \chi_{0}(q) + \chi_{1/2}(q) \right) \label{eq:Z_N}
\end{equation}

The spectrum with two Neumann defect insertions is identical to that of two corresponding Dirichlet defects with the fusion of two Kramers-Wannier defects:
\begin{equation}
Z^{s,t}_{N(\phi_1),N(\phi_2)}(q)
= Z^{s,0}_{D(\phi_1),D(\phi_2)}(q) + Z^{s,1}_{D(\phi_1),D(\phi_2)}(q)
\end{equation}

For fermionic theories, the partition function expressions $Z^{s, t}$ are the same but with the interpretation that $s+t$ labels the boundary condition (0 for Neveu-Schwarz, 1 for Ramond) and $s$ labels the global fermion parity $(-1)^F$ charge \cite{Karch:2019lnn,Ji:2019ugf,Hsieh:2020uwb,Fukusumi:2021zme}. This explains why $H_F^{\mu}$ and $H_F^{\mu}$ realize a stable nondegenerate interface with $H_F$ and a degenerate interface with each other.

\begin{center}
\small \textbf{Defect flows}
\end{center}

\begin{figure}[h!]
\centering\includegraphics[width=0.8\linewidth]{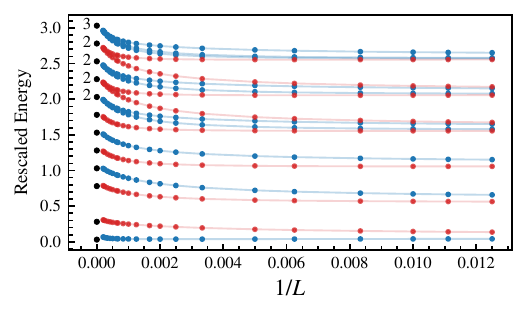}
    \caption{\textbf{Defect renormalization-group flow} A Neumann defect with Affleck–Ludwig entropy $g=\sqrt{2}$ flows under its most relevant defect perturbation to a lower-$g$ fixed point. For the $H_{\rm Ising}^{\sigma}|H_{\rm Ising}$ interface, symmetry forces the IR fixed point to be the factorizing $(\mathrm{free},\mathrm{free})$ defect. We confirm this numerically: the defect–endpoint spectrum (single defect insertion with $h=0.1$, $b=1$) splits into odd (red) and even (blue) fermion–parity sectors and, as $L\!\to\!\infty$, converges to the known conformal spectrum of the $\ket{f}\bra{f}$ defect (black dots), consistent with the phase diagram in Fig.~\ref{fig:Ising_interface}. 
    For the $H_{\rm Ising}^{\mu}| H_{\rm Ising}$ interface, the blue and red energy levels can be viewed as spectra for two-fold degenerate spontaneously aligned and anti‐aligned defect insertions.}
        \label{fig:N_to_ff}
\end{figure}

Defect flows from $N(\phi)$ to generic Dirichlet defects were studied in detail in Ref.~\cite{Kormos_2009, Fendley_2009}. At the $H^\sigma_{\rm Ising}$ to $H_{\rm Ising}$ interface, 
symmetry pins the fate of the Dirichlet defect to be $\ket{f}\bra{f}$. This is what occurs for any $b$ with $h \neq 0$ in Eq.~\ref{eq:Hs_H}. We show numerical data of the energy spectrum of a single defect insertion at $b=1$ and $h=0.1$ in Fig.~\ref{fig:N_to_ff}. The energies $E_j$ are shown rescaled to extract endpoint scaling dimensions, $E_j =- \frac{4}{\pi}L + \textrm{const.}+ \frac{4\pi}{L}(\Delta_j-c/12) $. While at small size the spectrum resembles that of the unperturbed Neumann defect (Eq.~\ref{eq:Z_N}), in the large $L$ limit it matches that of the $\ket{f}\bra{f}$ boundary condition $\chi_0(q^{1/2})+\chi_{1/2}(q^{1/2})$. 

The same analysis can be done of the defect endpoint spectrum for the perturbed interface between $H_{\rm Ising}$ and $H_{\rm Ising}^\mu$; in that case the plot in Fig.~\ref{fig:N_to_ff} looks identical but with the interpretation that for $h >0$ ( $h <0$) we see the blue (or red) data points with exact two-fold degeneracy, corresponding to the endpoint spectrum of the spontaneously aligned (or anti-aligned) defect $2\chi_0(q^{1/2})$ (or $2\chi_{1/2}(q^{1/2})$).
A version of this interface appears in Eq.~(4.1) of Ref.~\onlinecite{Grimm_1990} up to unitary equivalence and an added $\mathbb{Z}_2$-breaking Hamiltonian term; despite the latter they recover the same scaling dimensions of the defect Hilbert space's ground state.

\subsection*{Dictionary of Ising boundary operators}

In tables \ref{tab:dict-Hsigma-H}, \ref{tab:dict-H-Hmu} and \ref{tab:dict-Hsigma-Hmu}, for each $\mathbb{Z}_2 \times \mathbb{Z}_2^T$-symmetric Ising interface setup appearing in the main text realizing factorizing defects, we list the lattice operators realizing boundary CFT fields and their dimensions $\Delta$.
Boundary operators for boundary condition $B$ are denoted $\varphi^{B}$ (with $\mathbb{I}$ being the identity operator on $B$), while boundary-condition-changing operators from $B$ to $B'$ are denoted $\mu^{BB'}$. The subscripts $L/R$ refer to left and right boundaries; for simplicity we drop the subscripts when the boundaries differ.

The notation is chosen such that the unitary $\mathbb{Z}_2$ symmetry exchanges $\uparrow$ and $\downarrow$. 
On the spontaneously-fixed boundary of $H^\sigma_{\rm Ising}$, the anti-unitary time-reversal also acts by exchanging $\uparrow$ and $\downarrow$. 
On the free boundary of $H^\sigma_{\rm Ising}$, the time-reversal sends $\sigma \to - \sigma$. On the spontaneously-fixed boundary of $H^\mu_{\rm Ising}$, anti-unitary time-reversal sends $\mu^{\uparrow \downarrow} \to - \mu^{\uparrow \downarrow}$ and $\mu^{\downarrow \uparrow} \to - \mu^{\downarrow \uparrow}$.

%Table for Interface 1 H sigma to H 
\begin{table}[t]
\centering
\footnotesize
\setlength{\tabcolsep}{3pt}      % <-- shrink column padding
\renewcommand{\arraystretch}{1.15}

\begin{minipage}[t]{0.49\columnwidth}
\centering
\textbf{$|f\rangle\langle f|\quad (h,b)=(1,0)$}\par\smallskip
\begin{tabular}{@{}llc@{}}
\toprule
Op. & Field & $\Delta$ \\
\midrule
$X_{-1},X_{0}$ & $\mathbb{I}^{(f)}$ & $0$ \\
$Y_{-1}$ & $\sigma^{(f)}_{L}$ & $\tfrac12$ \\
$Z_{-1}$ & $\partial_t\sigma^{(f)}_{L}$ & $\tfrac32$ \\
$Z_{0}$  & $\sigma^{(f)}_{R}$ & $\tfrac12$ \\
$Y_{0}$  & $\partial_t\sigma^{(f)}_{R}$ & $\tfrac32$ \\
\bottomrule
\end{tabular}
\end{minipage}
\hfill
\begin{minipage}[t]{0.49\columnwidth}
\centering
$|f\rangle(\bra{\uparrow}+\bra{\downarrow})\quad (h,b)=(0,0)$\par\smallskip
\begin{tabular}{@{}llc@{}}
\toprule
Op. & Field & $\Delta$ \\
\midrule
$X_{-1}$ & $\mathbb{I}^{(f)}$ & $0$ \\
$Y_{-1}$ & $\sigma^{(f)}$ & $\tfrac12$ \\
$Z_{-1}$ & $\partial_t\sigma^{(f)}$ & $\tfrac32$ \\
$Z_{0}$  & $\mathbb{I}^{\uparrow}-\mathbb{I}^{\downarrow}$ & $\tfrac12$ \\
$X_{0}$  & $\mu^{\uparrow\downarrow}+\mu^{\downarrow\uparrow}$ & $\tfrac12$ \\
$Y_{0}$  & $i(\mu^{\uparrow\downarrow}-\mu^{\downarrow\uparrow})$ & $\tfrac12$ \\
\bottomrule
\end{tabular}
\end{minipage}

\caption{Dictionary between the lattice and the continuum for the boundary operators on the $H^\sigma_{\rm Ising} | H_{\rm Ising}$ interface (see Eq.~\eqref{eq:Hs_H_detail}).}
\label{tab:dict-Hsigma-H}
\end{table}

\begin{table}[t]
\centering
\footnotesize
\setlength{\tabcolsep}{3pt}
\renewcommand{\arraystretch}{1.15}

\begin{minipage}[t]{0.49\columnwidth}
\centering
$\ket{\uparrow} \bra{\uparrow} + \ket{\downarrow} \bra{\downarrow} $\quad $(h,b)=(1,0)$\par\smallskip
\begin{tabular}{@{}llc@{}}
\toprule
Op. & Field & $\Delta$ \\
\midrule
$Z_{0}$     & $\mathbb{I}_L^{\uparrow}\mathbb{I}_R^{\uparrow}-\mathbb{I}_L^{\downarrow}\mathbb{I}_R^{\downarrow}$ & $0$ \\
$X_{0}Z_{1}$& $i(\mu_L^{\uparrow\downarrow}\mu_R^{\uparrow\downarrow}-\mu_L^{\downarrow\uparrow}\mu_R^{\downarrow\uparrow})$ & $1$ \\
$Y_{0}Z_{1}$& $\mu_L^{\uparrow\downarrow}\mu_R^{\uparrow\downarrow}+\mu_L^{\downarrow\uparrow}\mu_R^{\downarrow\uparrow}$ & $1$ \\
$X_{0}$     & $\partial_t(\mu_L^{\uparrow\downarrow}\mu_R^{\uparrow\downarrow}+\mu_L^{\downarrow\uparrow}\mu_R^{\downarrow\uparrow})$ & $2$ \\
$Y_{0}$     & $i\partial_t(\mu_L^{\uparrow\downarrow}\mu_R^{\uparrow\downarrow}-\mu_L^{\downarrow\uparrow}\mu_R^{\downarrow\uparrow})$ & $2$ \\
\bottomrule
\end{tabular}
\end{minipage}
\hfill
\begin{minipage}[t]{0.49\columnwidth}
\centering
$|f\rangle(\bra{\uparrow}+\bra{\downarrow})\quad (h,b)=(0,0)$
\par\smallskip
\begin{tabular}{@{}llc@{}}
\toprule
Op. & Field & $\Delta$ \\
\midrule
$X_{-1}$     & $\mathbb{I}$ & $0$ \\
$Z_{-1}$     & $\sigma^{(f)}$ & $\tfrac12$ \\
$Y_{-1}$     & $\partial_t\sigma^{(f)}$ & $\tfrac32$ \\
$Z_{0}$      & $\mathbb{I}^{\uparrow}-\mathbb{I}^{\downarrow}$ & $0$ \\
$X_{0}Z_{1}$ & $i(\mu^{\uparrow\downarrow}-\mu^{\downarrow\uparrow})$ & $1$ \\
$Y_{0}Z_{1}$ & $\mu^{\uparrow\downarrow}+\mu^{\downarrow\uparrow}$ & $1$ \\
\bottomrule
\end{tabular}
\end{minipage}

\caption{Dictionary between the lattice and the continuum for the boundary operators on the $H_{\rm Ising}|H^\mu_{\rm Ising}$ interface (see Eq.~\eqref{eq:H_Hu_detail}). 
Here on the $\ket{\uparrow} \bra{\uparrow} +\ket{\downarrow}\bra{\downarrow} $ defect, one should note that the individual factors $\mu_L$ and $\mu_R$ do not exist as point  operators on the interface.} 
\label{tab:dict-H-Hmu}
\end{table}

\begin{table}[t]
\centering
\footnotesize
\setlength{\tabcolsep}{3pt}
\renewcommand{\arraystretch}{1.15}

\begin{minipage}[t]{0.49\columnwidth}
\centering
$|f\rangle(\bra{\uparrow}+\bra{\downarrow})\quad (h,\theta)=(1,0)$\par\smallskip
\begin{tabular}{@{}llc@{}}
\toprule
Op. & Field & $\Delta$ \\
\midrule
$X_{0}$ & $\mathbb{I}$ & $0$ \\
$Y_{0}$ & $\sigma^{(f)}$ & $\tfrac12$ \\
$Z_{0}$ & $\partial_t\sigma^{(f)}$ & $\tfrac32$ \\
$Z_{1}$ & $\mathbb{I}^{\uparrow}-\mathbb{I}^{\downarrow}$ & $0$ \\
$X_{1}Z_{2}$ & $i(\mu^{\uparrow\downarrow}-\mu^{\downarrow\uparrow})$ & $\tfrac12$ \\
$Y_{1}Z_{2}$ & $\mu^{\uparrow\downarrow}+\mu^{\downarrow\uparrow}$ & $\tfrac12$ \\
\bottomrule
\end{tabular}
\end{minipage}
\hfill
\begin{minipage}[t]{0.49\columnwidth}
\centering
$(\ket\uparrow+\ket\downarrow)(\bra{\uparrow}+\bra{\downarrow})\quad h=0$\par\smallskip
\begin{tabular}{@{}llc@{}}
\toprule
Op. & Field & $\Delta$ \\
\midrule
$Y_{0}$ & $\mathbb{I}_L^{\uparrow}-\mathbb{I}_L^{\downarrow}$ & $0$ \\
$X_{0}$ & $\mu_L^{\uparrow\downarrow}+\mu_L^{\downarrow\uparrow}$ & $\tfrac12$ \\
$Z_{0}$ & $i(\mu_L^{\uparrow\downarrow}-\mu_L^{\downarrow\uparrow})$ & $\tfrac12$ \\
$Z_{1}$ & $\mathbb{I}_R^{\uparrow}-\mathbb{I}_R^{\downarrow}$ & $0$ \\
$X_{1}Z_{2}$ & $i(\mu_R^{\uparrow\downarrow}-\mu_R^{\downarrow\uparrow})$ & $\tfrac12$ \\
$Y_{1}Z_{2}$ & $\mu_R^{\uparrow\downarrow}+\mu_R^{\downarrow\uparrow}$ & $\tfrac12$ \\
\bottomrule
\end{tabular}
\end{minipage}

\caption{Dictionary between the lattice and the continuum for the boundary operators on the $H^\sigma_{\rm Ising} |H^\mu_{\rm Ising}$ interface at $h=1, \theta =0$ (see Eq.~\eqref{eq:Hs_Hmu_detail}). }
\label{tab:dict-Hsigma-Hmu}
\end{table}

\section{Degeneracy of the $H^{\sigma}_{\rm Ising}|H^{\mu}_{\rm Ising}$ interface}\label{app:Hsimga_Hmu_defects_proof}

Here we show that all $g<\sqrt{2}$ defects and all non-degenerate defects are forbidden by the symmetry between $H^\sigma_\text{Ising}$ and $H^\mu_\text{Ising}$. 

First, following the same argument at the end of Sec.~\ref{subsec:Hs_H}, the vanishing two-point functions of the $\sigma$ field across the interface forbids all the Dirichlet defects except for $\ket{f}\bra{f}$. Furthermore, note that   $H^\sigma_\text{Ising}$ and $H^\mu_\text{Ising}$ are related by the lattice Kramers-Wannier transformation, which maps $\ket{f}\bra{f}$ to a spontaneously aligned or anti-aligned defect \cite{Fukusumi:2021zme}. 
However, since the latter are forbidden by the two-point function argument above, this means that even the $\ket{f}\bra{f}$ defect is not allowed.
Thus, all $g=1$ defects are forbidden.

The Neumann defects all have vanishing two-point function of $\sigma$, but we can gain additional constraints by using the defect anomaly arguments of section \ref{sec:projective}. First, we can apply an entangler on the left side $\prod_{j\ge0} \mathrm{CZ}_{j,j+1}$ followed by a basis rotation on the right side $\prod_{j\le 0} e^{i\pi X_j/4}$ to turn both sides of the chain to asymptotically the standard Ising Hamiltonian $H_{\rm Ising}$. Crucially this local unitary transformation  sends the original symmetry generators of $\mathbb{Z}_2 \times \mathbb{Z}_2^T$ to
\[
\prod_j X_j,
\qquad
\mathcal{K}\, Z_0 \prod_{j\le -1} X_j .
\]

These symmetry generators cannot be placed on a ring with periodic boundary conditions. However, we can fuse a  Kramers-Wannier topological defect $N(\pi/4)$ to the left of this interface at site $-1$ through the standard semi-infinite operation \cite{Thorngren:2021yso}
\[ X_{j\le -1} \to Z_{j-1} Z_j \quad Z_{j\le -1} \to \prod_{k=j}^{-1} X_k\]
converting the symmetry generators for this fused defect to
\[
Z_{-1} \prod_j X_j \qquad \mathcal{K} Z_{-1} Z_0 
\]
These generators realize a projective representation on periodic boundary conditions. Thus the fused defect  
${N}(\pi/4)\cdot {\cal D}$ 
has a defect 't Hooft anomaly. 

Suppose there was a marginal deformation from $N(0) = \ket{f}\bra{ \uparrow} + \ket{f}\bra{ \downarrow}$ to a non-degenerate $N(\phi)$ with $0<\phi<\pi/2$. Then $N(\pi/4)\cdot  N(\phi) = D(\phi)+D(\pi-\phi)$
, which does \textit{not} have endpoint degeneracy. Thus, such marginal deformations are incompatible with symmetries.

We conclude that the $H^\sigma _\text{Ising}| H_\text{Ising}^\mu$ interface must, without exception, decompose into a direct sum of interfaces and exhibit degenerate edge modes.

We comment on the interfaces presented in Fig~\ref{fig:Ising_interface}. Fine-tuned, degenerate defects are allowed and generally unstable. For example $h =0$ in Eq.~\eqref{eq:Hs_Hmu_detail} realizes the $D(0)+D(\pi) = (\ket{\uparrow} +\ket{ \downarrow})(\bra{ \uparrow}+\bra{\downarrow})$ interface, whose two symmetric relevant operators $\mu^{\uparrow \downarrow}_L+\mu^{\uparrow \downarrow}_L$ and $i(\mu^{\uparrow \downarrow}_L-\mu^{\uparrow \downarrow}_L)(\mathbb{I}^{\uparrow}_R-\mathbb{I}^{\downarrow}_R)$ (tuned by $h$) both drive flows into the stable $N(0)$ defect with no further relevant symmetric perturbations. 
In fact, a unitary $\exp(i\theta Y_0 Z_1/2)$ tunes shifts in $\theta$, suggesting potential exploration of winding around a noncontractible loop of the $N(0)$ defect in the phase diagram and whether there are analogs to Thouless pumps. 

As an aside, 
folding the $H_F^\sigma|H_F^\mu$ interface can be implemented by the relabeling (for $j\ge 0$)
\[
\begin{array}{rcl@{\qquad}rcl}
\gamma_{2j} &\mapsto& \gamma_{4j}        & \gamma_{2j+1}     &\mapsto& \gamma_{4j+3}\\
\gamma_{-(2j+1)} &\mapsto& \gamma_{4j+1} & \gamma_{-(2j+2)}  &\mapsto& \gamma_{4j+2}
\end{array}
\]
which rewrites the interface as an open boundary of a $c=1$ Majorana chain,
\[
H^{\text{fold}}_F \;=\; i\sum_{j\ge 0}\!\big(\gamma_{2j+1}\gamma_{2j+2} + \gamma_{2j+1}\gamma_{2j+6}\big).
\]
With any symmetric boundary interactions, this folded model supports robust, symmetry-protected degenerate edge modes.

A Jordan-Wigner transformation then leads to a convenient spin-chain description (we refer to it as the \emph{symmetry-enriched XX chain})

\begin{equation}
    H_{\rm XX} \;=\; -\sum_j \!\left( Z_j Z_{j+1} \;+\; Z_{j-1} X_j X_{j+1} Z_{j+2} \right)
    \label{eq:sym-enriched-dqcp}
\end{equation}

This spin chain has the global $\mathbb{Z}_2$ symmetry $U_X=\prod_j X_j$ and complex conjugation $\mathcal{K}$. On an open chain, generic symmetry-allowed boundary perturbations lift the edge degeneracy with a finite-size splitting $\Delta E \propto L^{-3}$

If the protecting symmetry is enhanced to $\mathbb{Z}_2\times \mathbb{Z}_2 \times \mathbb{Z}_2^T$ by an additional global generator $U_Z=\prod_j Z_j$, the leading symmetry-allowed boundary perturbation responsible for the $L^{-3}$ splitting is forbidden.
Consequently, the edge-mode's algebraic localization is sharpened to 
\( \Delta E \propto 1/L^6\).
This conclusion follows from the boundary-operator analysis in
Sec.~\ref{subsec:Hs_Hmu_interface} for the Ising interface discussed there with a different gapped 
$\mathbb{Z}_2$ sector on either side.

\begin{figure}
  \centering
  \includegraphics[width=0.5\linewidth]{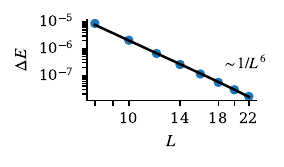}
  \caption{\textbf{Finite-size energy splitting of edge modes at the symmetry-enriched DQCP.}
Exact-diagonalization data for the open-chain Hamiltonian $H_{\rm XX}^{\rm bdy}$ in
Eq.~\eqref{eq:sym-enriched-dqcp-bdy} ($L=2N$, $b=0.1$), which preserves the symmetry generated by
$U_X=\prod_j X_j$, $U_Z=\prod_j Z_j$, and complex conjugation $\mathcal K$.
The difference in energies of the two symmetry-protected edge-modes scales as $\Delta E\sim L^{-6}$.}
  \label{fig:gap_L6}
\end{figure}

In Fig.~\ref{fig:gap_L6} we illustrate this behavior by exact diagonalization of the open-chain Hamiltonian
\begin{align}
H_{\rm XX}^{\rm bdy} &=-\sum_{j=1}^{2N-1} Z_j Z_{j+1}
  -\sum_{j=1}^{2N-3} Z_j X_{j+1} X_{j+2} Z_{j+3}\notag\\
  &
  - b( X_1 X_2 + X_{2N-1}X_{2N}) 
  \label{eq:sym-enriched-dqcp-bdy}
\end{align}
with $L=2N$ and (for the data shown) $b=0.1$.
With this enhanced symmetry, the model realizes a symmetry-enriched deconfined quantum critical point (DQCP) transitioning
between $U_X$-preserving and $U_Z$-preserving symmetry-broken orders. A different enrichment of the same SSB transition
appears in Ref.~\onlinecite{Rey_2025}.

\end{document}